\documentclass[aps,prd,preprint,superscriptaddress,nofootinbib]{revtex4-1}
\usepackage{braket}
\usepackage{bm}
\usepackage{xargs}
\usepackage{mathtools}
\usepackage[english]{babel}
\usepackage[utf8]{inputenc}
\usepackage{amsmath,amssymb,braket,slashed,mathrsfs,bm}
\usepackage{pifont}
\usepackage{graphicx}
\usepackage{tikz}
\usetikzlibrary{shapes,arrows,positioning}
\tikzset{decision/.style={diamond, draw, fill=blue!20, text width=4.5em, text badly centered, inner sep=0pt}}
\tikzset{block/.style={rectangle, draw, fill=blue!20, text width=10em, text centered, rounded corners, minimum width=3.5cm}}
\tikzset{block1/.style={rectangle, draw, fill=blue!20, text width=18.5em, text centered, rounded corners, minimum width=3.5cm}}
\tikzset{line/.style={draw, -latex, thick}}
\usepackage{color,hyperref}
\usepackage{multirow,array}

\usepackage{cleveref}
\newcommand{\ba}{\begin{eqnarray}}
\newcommand{\ea}{\end{eqnarray}}
\newcommand{\be}{\begin{equation}}
\newcommand{\ee}{\end{equation}}
\newcommand{\bi}{\begin{itemize}}
\newcommand{\ei}{\end{itemize}}

\newcommand{\nn}{\nonumber}

\newcommand{\dfour}{d^{\hspace{1pt}4}}
\newcommand{\dthree}{d^{\hspace{1pt}3}}
\newcommand{\vs}{\hspace{1pt}}
\newcommand{\pslash}{\slash{\hspace{-6.6pt}p}}

\newcommand{\cu}{Physics Department, Columbia University, New York, NY 10027, USA}
\newcommand{\soton}{Department of Physics, University of Southampton, Southampton SO17 1BJ, UK}
\newcommand{\innovation}{Collaborative Innovation Center of Quantum Matter, Beijing 100871, China}
\newcommand{\chep}{Center for High Energy Physics, Peking University, Beijing 100871, China}
\newcommand{\pkuphy}{School of Physics, Peking University, Beijing 100871,
China}
\newcommand{\KeyLab}{State Key Laboratory of Nuclear Physics and Technology,
Peking University, Beijing 100871, China}
\newcommand{\UCONN}{Physics Department, University of Connecticut, Storrs, Connecticut 06269-3046, USA}
\newcommand{\RBRC}{RIKEN BNL Research Center, Brookhaven National Laboratory, Upton, New York 11973, USA}

\begin{document}
\title{Finite-volume effects in long-distance processes with massless leptonic propagators}

\author{Norman~H.~Christ}\email{nhc@phys.columbia.edu}\affiliation{\cu}
\author{Xu~Feng}\email{xu.feng@pku.edu.cn}\affiliation{\pkuphy}\affiliation{\innovation}\affiliation{\chep}\affiliation{\KeyLab}
\author{Lu-Chang~Jin}\email{ljin.luchang@gmail.com}\affiliation{\UCONN}\affiliation{\RBRC}
\author{Christopher~T.~Sachrajda}\email{cts@soton.ac.uk}\affiliation{\soton}
\pacs{PACS}

\date{\today}

\begin{abstract}
In Ref.\,\cite{Feng:2018qpx}, a method was proposed to calculate QED corrections to hadronic self energies from lattice QCD without power-law finite-volume errors. In this paper, we extend the method to processes which occur at second-order in the weak interaction and in which there is a massless (or almost massless) leptonic propagator. We demonstrate that, in spite of the presence of the propagator of an almost massless electron, such an \emph{infinite-volume reconstruction} procedure can be used to obtain the amplitude for the rare kaon decay $K^+\to\pi^+\nu\bar\nu$ from a lattice quantum chromodynamics computation with only exponentially small finite-volume corrections. 
\end{abstract}

\maketitle
   
\section{Introduction}

Lattice Quantum Chromodynamics (QCD) has been successful in precision flavor
physics, where observables such as the
decay constants $f_K$, $f_\pi$ and semileptonic form factors $f_+(0)$ can be calculated with sub-percent precision\,\cite{Aoki:2019cca}. The quantities mentioned above, which occur at leading order in the weak interaction, provide important constraints for CKM matrix elements. With the development of supercomputers, algorithms and new ideas, the range of lattice QCD
calculations has been extended to include many second-order electroweak processes, where the calculations involve
the construction of 4-point correlation functions and the treatment of bilocal 
matrix elements with the insertion of two operators from the effective Hamiltonian. Examples include
$K_L$-$K_S$ mixing\,\cite{Christ:2012se,Bai:2014cva,Christ:2015phf,Wang:2018csg} and $\epsilon_K$\,\cite{Christ:2015phf}, 
rare kaon
decays\,\cite{Feng:2015kfa,Christ:2016eae,Christ:2016lro,Christ:2016psm,Bai:2017fkh,Bai:2018hqu,Christ:2019dxu,Christ:2015aha,Christ:2016awg,Christ:2016mmq,Lawson:2017kxc},
neutrinoless double-beta
decays\,\cite{Tiburzi:2017iux,Shanahan:2017bgi,Nicholson:2018mwc,Feng:2018pdq,Nicholson:2018laj,Tuo:2019bue,Cirigliano:2020yhp,Detmold:2020jqv},
electroweak box contributions to semileptonic decays~\cite{Feng:2020zdc},
inclusive $B$-meson decays\,\cite{Hashimoto:2017wqo,Hashimoto:2019pgh,Gambino:2020crt}, nucleon
Compton amplitudes\,\cite{Chambers:2017dov,Liu:2017lpe,Hansen:2017mnd,Can:2020sxc}, as well
as isospin-breaking effects in hadronic
spectra\,\cite{Duncan:1996xy,Duncan:1996be,Blum:2007cy,Blum:2010ym,Ishikawa:2012ix,deDivitiis:2013xla,Borsanyi:2014jba,Horsley:2015vla,Giusti:2017dmp,Davoudi:2018qpl},
the hadronic vacuum polarization
function\,\cite{Giusti:2017jof,Chakraborty:2017tqp,Boyle:2017gzv,Blum:2018mom},
leptonic decays\,\cite{Carrasco:2015xwa,Lubicz:2016xro,Giusti:2017dwk} and 
$K\to\pi\pi$ decays\,\cite{Christ:2017pze,Cai:2018why}.

When analyzing matrix elements of bilocal operators, it is useful
to insert a complete set of intermediate states between the two local operators. If the energy of the initial state is sufficiently
large to create on-shell intermediate
multi-particle states, power-law finite-volume effects can be
generated\,\footnote{Throughout this paper we use the shorthand notation \emph{exponential finite-volume effects} to denote ones which decrease exponentially with the spacial extent of the lattice $L$, and \emph{power-law finite-volume effects} to denote those which decrease only as powers of $L$.}. Following Ref.\,\cite{Christ:2015pwa}, where the
$K_L$-$K_S$ mixing is analysed as an example, one can correct for such potentially large
finite-volume effects. However, the situation changes when
the intermediate multi-particle state involves a massless, or nearly massless, particle. 
Since the long-range massless propagator is distorted by the finite volume, 
power-law finite-volume effects appear even for states containing off-shell particles. Such a situation happens, for example, in the rare kaon decay 
$K^+\to X
e^+\nu_e\to\pi^+\nu_e\bar{\nu}_e$\,\cite{Feng:2015kfa,Christ:2016eae,Christ:2016psm,Christ:2016lro,Bai:2017fkh,Bai:2018hqu}
where the intermediate states contain a positron,
together possibly with additional hadronic particles specified here by the symbol $X$. The positron is effectively massless since its mass, $m_e$, satisfies $m_eL\ll 1$, where $L$ is the spacial extent of current lattices (with volume, $V=L^3$).  An analogous procedure has been applied to the calculation of the amplitude for neutrinoless double-$\beta$ decay $\pi^-\pi^-\to X e\bar{\nu}_e \to
ee$, in which there is the propagator of a massless neutrino\,\cite{Feng:2018pdq}.

To completely remove the power-law finite-volume effects induced by the
massless electron or neutrino, we adopt the {\em infinite-volume reconstruction} (IVR) 
method proposed in Ref.\,\cite{Feng:2018qpx}, which has been used to eliminate such effects in QED corrections to hadronic self energies. In that case the lightest intermediate hadron is the same as the stable hadron in the initial and final
states. In this paper, we use the rare decay $K^+\to\pi^+\nu\bar{\nu}_\ell$ to illustrate that the method is also applicable to processes in which
the intermediate hadronic state is not degenerate with the initial state; indeed it can be either heavier or lighter than initial state.

The structure of the remainder of this paper is as follows. In the following section we discuss the structure and properties of the physical amplitude for the rare kaon decay $K^+\to\pi^+\nu\bar\nu$ in Minkowski space and write the amplitude in a form which is convenient for continuation into Euclidean space. In Sec.\,\ref{sec:Euclidean} we present our proposed method for the evaluation of the amplitude from Euclidean correlation functions computed in a finite-volume, but with only exponentially small finite-volume corrections. Finally we present our conclusions in Sec.\,\ref{sec:concs}.

\section{Finite-volume effects in {\boldmath$K^+\to\pi^+\nu_\ell\bar{\nu}_\ell$} decays}\label{sec:Minkowski}

\begin{figure}
\begin{center}        
\includegraphics[width=.4\hsize]{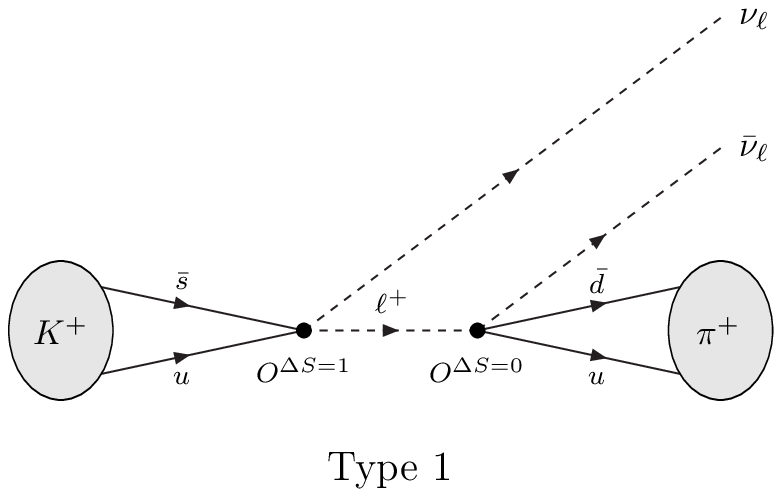}\hspace{0.5in}
\includegraphics[width=.4\textwidth]{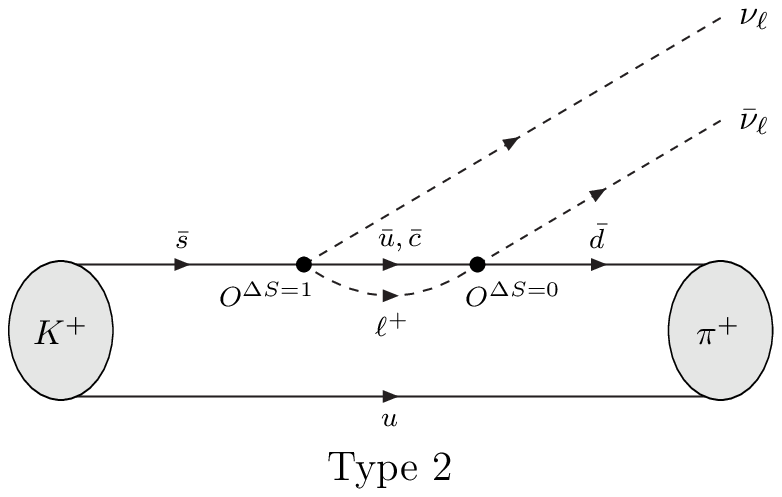}
 \caption{Quark and lepton contractions for the $W$-$W$ diagrams. The quark flavors are as indicated and the lepton $\ell=e,\mu$ or $\tau$. The quark (leptons) are represented by solid (dashed) lines.The operators $O^{\Delta S=1}$ and $O^{\Delta S=0}$ are shorthand representations of $O_{q\ell}^{M,\Delta S=1}$ and $O_{q\ell}^{M,\Delta S=0}$ defined in Eq.\,(\ref{eq:OWW}).}
   \label{fig:Wbox}
   \end{center}
   \end{figure}

  As explained in Ref.\,\cite{Christ:2016eae}, the $K^+\to\pi^+\nu_\ell\bar{\nu}_\ell$ decay amplitude, where $\ell$ represents the lepton quantum number, contains contributions from both $Z$-exchange
   diagrams and $W$-$W$
   diagrams. For the $Z$-exchange diagrams, for which the $\nu\bar\nu$ pair is emitted from the same vertex, there are no leptonic propagators in the amplitude and the dominant, power-law finite-volume effects are associated with the 
  process $K^+\to\pi^+\pi^0\to\pi^+\nu_\ell\bar{\nu}_\ell$, which can be corrected using the formula provided in Ref.\,\cite{Christ:2015pwa}. Here we focus on the contribution from the $W$-$W$
   diagrams illustrated in Fig.\,\ref{fig:Wbox} in which the $\nu_\ell$ and $\bar\nu_\ell$ are emitted from separate vertices and which contain the propagator of the corresponding charged lepton $\ell^+\,(e^+,\mu^+$ or $\tau^+)$. The discussion of the properties and structure of the physical amplitude in this section is presented in Minkowski space.  

\label{subsubsec:scalar}

   The contribution to the amplitude 
   from the $W$-$W$ diagrams in Minkowski space, $A^{M}_\ell$, is given by
   \be
   A^{M}_\ell=A_{u,\ell}^M-A_{c,\ell}^M
   \ee
   where the $A_{q,\ell}^{M}$ are defined by
   \begin{eqnarray}\label{eq:AqMdef}
       A_{q,\ell}^{M}=i\int \dfour x~\langle\,\pi^+\nu_\ell\bar{\nu}_\ell\,|\,T\big\{O_{q\ell}^{M,\Delta S=1}(x)\,
   O_{q\ell}^{M,\Delta S=0}(0)\big\}\,|\,K^+\,\rangle\,,
   \end{eqnarray}
where $q = u, c$ are the flavors of up-type quarks and $\ell = e,\mu,\tau$ are the flavors of leptons.
The two operators in Eq.\,(\ref{eq:AqMdef}) are given by
   \begin{equation}
   O_{q\ell}^{M,\Delta S=1}=(\bar{s}q)_{V-A}(\bar{\nu}_\ell\ell)_{V-A},\quad
   O_{q\ell}^{M,\Delta S=0}=(\bar{q}d)_{V-A}(\bar{\ell}\nu_\ell)_{V-A}\,,\label{eq:OWW}
   \end{equation}
where, for example, $(\bar{s}q)_{V-A}(\bar{\nu}_\ell\ell)_{V-A}\equiv \big(\bar{s}\gamma^\mu(1-\gamma^5)q\big)
\big(\bar{\nu}_\ell\gamma_\mu(1-\gamma^5)\ell\big)$.
    
   In this paper we focus on the transition
   $K^+\to X\ell^+\nu_\ell\to\pi^+\nu_\ell\bar{\nu}_\ell$.
   We denote the potentially large, i.e.~the
   power-law, finite-volume effects in the spatial integral over a finite volume of size $L^3$, by $A_{\textrm{FV}}^{X\ell^+}=A^{X\ell^+}(L)-A^{X\ell^+}(\infty)$, where $A^{X\ell^+}(L)$
   and $A^{X\ell^+}(\infty)$ are the amplitudes in finite and infinite 
   volumes respectively. The label 
   {\footnotesize $X\ell^+$} indicates that the correction comes from the $X\ell^+$ intermediate states, where $X$ can represent (i) the vacuum, (ii) the stable single-hadron states, $\pi^0$ or $D^0$, or (iii)
multi-hadron states of which the lightest ones are two-pion states. The neutrino in the $X\ell^+\nu_\ell$ intermediate state is the one which appears in the final state, and its energy and momentum determine those of $X\ell^+$.
   
For the transition $K^+\to Y\ell^-\bar{\nu}_\ell\to\pi^+\nu_\ell\bar{\nu}_\ell$, charge conservation requires $Y$ to be a multi-hadron state. The corresponding finite-volume effects are largely similar to those from multi-hadron states 
$X$ in $K^+\to X\ell^+\nu_\ell\to\pi^+\nu_\ell\bar{\nu}_\ell$ transitions apart from the presence of \emph{disconnected} diagrams as discussed in Sec.\,\ref{subsec:tplus}.

We now consider the three possibilities for $X$ in turn. When $X$ is the vacuum, the momentum of $\ell^+$ is completely fixed by  momentum conservation, $p_{\ell^+}=p_K-p_{\nu_\ell}\equiv P$. There are no power-law finite-volume effects in this case.

   \begin{figure}
   \centering
        \includegraphics[width=.45\textwidth]{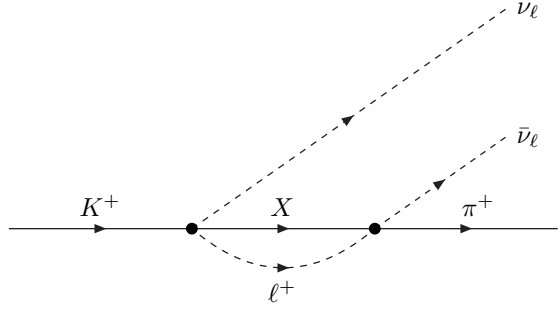}
       \caption{Illustration of the process $K^+\to X\ell^+\nu_\ell\to
       \pi^+\nu_\ell\bar{\nu}_\ell$.}
   \label{fig:X_contribution}
   \end{figure}

   When $X$ is a single stable hadron with four momentum $k$, as shown in Fig.~\ref{fig:X_contribution}, the finite-volume effects in the amplitude, $A_{\textrm{FV}}^{X\ell^+}$, can be expressed as\,\cite{Christ:2016eae,Bai:2018hqu}
   \ba
   \label{eq:WW_FV}
   A_{\textrm{FV}}^{X\ell^+}&=&\left(\frac{1}{L^3}\sum_{\vec{k}}\int \frac{dk_0}{2\pi}
   -\int\frac{\dfour k}{(2\pi)^4}\right)
   \left\{A_\alpha^{K^+\to X}(p_K,k)\frac{i}{k^2-m_X^2+i\varepsilon}
    A_\beta^{X\to\,\pi^+}(k,p_\pi)\right\}
   \nn\\
   &&\hspace{1.5cm}\times
   \left\{\bar{u}(p_{\nu_\ell})\gamma^\alpha(1-\gamma_5)
   \frac{i}{({\slashed P}-{\slashed k})-m_{\ell}+i\varepsilon}
   \gamma^\beta(1-\gamma_5)v(p_{\bar{\nu}_\ell})\right\},
   \ea
   where $k$ is the momentum carried by the intermediate hadron $X$ and $P=p_K-p_{\nu_\ell}$ is the total
   momentum flowing into the $X\ell^+$ loop. The terms
   $A_\alpha^{K^+\to X}$ and $A_\beta^{X\to\,\pi^+}$ represent
   the transition matrix elements indicated by the superscripts and $\alpha,\,\beta$ are the Lorentz indices of the 
   weak currents.
   
Although the present study is focussed on rare kaon decays, the main ideas are more general. Equation\,(\ref{eq:WW_FV}) is an example of the generic form of the expression for finite-volume effects:
   \begin{eqnarray}
   \hspace{-0.25in}I_{\mathrm{FV}}=I(L)-I(\infty)=\left(\frac{1}{L^3}\sum_{\vec{k}}\int \frac{dk_0}{2\pi}
   -\int\frac{\dfour k}{(2\pi)^4}\right)\frac{f(k_0,{\bf
   k})}{(k^2-m_1^2+i\varepsilon)((P-k)^2-m_2^2+i\varepsilon)},
   \end{eqnarray}
   where, in the present calculation, $P=p_K-p_{\nu_\ell}\equiv(E,\vec{P})$, $m_1=m_X$ and $m_2=m_{\ell}$. 
 We can evaluate the $k_0$ integration using Cauchy’s
theorem, including the contributions from the poles in the
two propagators shown in Eq. (5) and ignoring contributions
from any other $k_0$ singularities in $f(k_0, k)$ since these
will result from other more massive intermediate states than
those we have chosen to study.  
For simplicity of notation, the dependence of $f(k_0,{\bf k})$
   on the external
   momenta is not shown explicitly. Performing the integral over $k_0$ we obtain the integrand
   \be\label{eq:twosingularities}
   -i\,\frac{f(E_{1},\vec{k})}{2E_{1}\big((E-E_{1})^2-E_{2}^2+i\varepsilon\big)}-i\,\frac{f(E+E_{2},\vec{k})}{2E_{2}\big((E+E_{2})^2-E_{1}^2+i\varepsilon\big)}
   \ee
  with $E_{1}=\sqrt{m_1^2+\vec{k}^2}$ and $E_{2}=\sqrt{m_2^2+(\vec{P}-\vec{k})^2}$. 
 Using the Poisson summation formula, it can be shown that 
 two singularities of the integrand contribute finite-volume effects which are not exponentially small in the volume.
 In the first term in Eq.\,(\ref{eq:twosingularities}) there is a 
  singularity when the condition $E=E_{1}+E_{2}$ is satisfied and two on-shell particles are created\,\cite{Kim:2005gf}. In the second term of Eq.\,(\ref{eq:twosingularities}), if $m_2$ is very small then there is an additional singularity from the factor $1/E_{2}\approx1/|\vec{P}-\vec{k}|$. 
For example, using a QED$_\mathrm{L}$-style regularization and omitting the
zero-momentum mode from the allowed finite-volume
lepton states, the region around $|\vec{P}-\vec{k}|=0$ leads to a $1/L^2$ difference between the finite-volume summation and infinite-volume integration\,\cite{Hasenfratz:1989pk,Lubicz:2016xro}.
This is the situation for rare kaon decays when the lepton is the electron where, since in practice $m_eL\ll 1$, the electron is effectively massless in lattice computations. Therefore, no matter how heavy is the hadron $X$,  power-law finite-volume effects associated with the massless electron always exist. In the following section we will discuss how to remove these two sources of  power-law finite-volume effects by extending the method developed in Ref.\,\cite{Feng:2018qpx}.
  
   When $X$ is a multi-hadron state, the situation is more complicated.
   For multi-hadron states with energies that are larger than the energy of the
   initial hadron, the contributions are exponentially
   suppressed at large time separations and the corresponding power-law
   finite-volume effects can be safely eliminated using our proposed method. For
   multi-hadron states with energies which are smaller than that of the
   initial hadron, it is unclear in general how to remove all the power-law
   finite-volume effects. Fortunately, for the $K^+\to\pi^+\nu_\ell\bar{\nu}_\ell$ decay, the contribution from low-lying multihadron states, e.g.
   $K^+\to2\pi\ell^+\nu_\ell\to\pi^+\nu_\ell\bar{\nu}_\ell$, can safely be neglected due to the significant phase space suppression.

\subsection{Structure of the amplitude}\label{subsec:structure}

Before explaining how to obtain the physical decay amplitude from a computation on a finite lattice we formulate it in an expression suitable for continuation into Euclidean space. We start by rewriting the bilocal matrix element in the integrand of $A^{M}_{q,\ell}$ in Eq.\,(\ref{eq:AqMdef}) as a product of two factors:
   \begin{equation}
   \label{eq:H_and_L}
       A_{q,\ell}^M=i\int d^{\hspace{1.5pt}4}x\,H^{M,(q) }_{\alpha\beta}(x)\,L^{M,\alpha\beta}(x)\,,
   \end{equation}
   where $\alpha$ and $\beta$ are Lorentz indices.
   The hadronic factor $H^{M,(q) }_{\alpha\beta}(x)$ and the leptonic factor  
   $L^{M,\alpha\beta}(x)$ are defined by
   \begin{eqnarray}
   H^{M,(q)}_{\alpha\beta}(x)&=&\langle\pi^+(p_\pi)|T\big\{\big[\bar{s}(x)\gamma_\alpha
   (1-\gamma_5)q(x)\big]\,\big[\bar{q}(0)\gamma_\beta(1-\gamma_5)d(0)\big]\,\big\}|K^+(p_K)\rangle
   \label{eq:HM}\\
   L^{M,\alpha\beta}(x)&=&\bar{u}(p_{\nu_\ell})\gamma^\alpha
   (1-\gamma_5)S_\ell(x,0)\gamma^\beta(1-\gamma_5){v}(p_{\bar{\nu}_\ell})\, e^{ip_{\nu_\ell} \cdot x}.\label{eq:LM}
   \end{eqnarray}
   Here $S_{\ell}(x,0)$ is a free lepton propagator.
By inserting a complete set of energy eigenstates, $A_{q,\ell}^{M}$ can further be written
as
\ba
A_{q,\ell}^{M}&=&\int d\phi_n\,
\frac{\langle \pi^+|O_{d,\beta}^{(q)}(0)|n\rangle\langle
n|O_{s,\alpha}^{(q)}(0)|K^+\rangle}{E_n+E_{\ell^+}+E_{\nu_\ell}-E_K-i\varepsilon}\,L_1^{\alpha\beta}(\vec{p}_n)
\nn\\
&&\hspace{0.2in}+\int d\phi_{n_s}
\frac{\langle \pi^+|O_{s,\alpha}^{(q)}(0)|n_s\rangle
\langle n_s|O_{d,\beta}^{(q)}(0)|K^+\rangle}{E_{n_s}+E_{\ell^-}+E_{\bar{\nu}_\ell}-E_K}
L_2^{\alpha\beta}(\vec{p}_{n_s})\label{eq:amplitudeM}
\\ &\equiv&A_{q,\ell}^{M,-}+A_{q,\ell}^{M,+}\,,\label{eq:amplitudeMmp}
\ea
where the first and second terms on the right-hand sides of Eqs.\,(\ref{eq:amplitudeM}) and (\ref{eq:amplitudeMmp}) are the contributions from the regions $x_0<0$ and $x_0>0$ respectively.
Here $|n\rangle$ and $|n_s\rangle$ represent non-strange and strangeness $S=1$ intermediate states respectively and in each case $\phi_n$ and $\phi_{n_s}$ is the corresponding phase space and a sum over all such states is implied. The denominator in the second term on the right-hand side of Eq.\,(\ref{eq:amplitudeM}) is always positive and hence we omit the $-i\varepsilon$.
The three-momenta of the charged leptons are fixed by the momentum-conserving $\delta$-functions obtained after the integration over $\vec{x}$, so that
\begin{eqnarray}
L_1^{\alpha\beta}(\vec{p}_n)&=& \frac{1}{2E_\ell^+} \,\bar{u}(p_{\nu_\ell})\gamma^\alpha(1-\gamma^5)(\pslash_{\ell^+}-m_\ell)
\gamma^\beta(1-\gamma^5)v(p_{\bar{\nu}_\ell})\,,\label{eq:L1}
\\
L_2^{\alpha\beta}(\vec{p}_{n_s})&=&\frac{1}{2E_\ell^-}\,\bar{u}(p_{\nu_\ell})\gamma^\alpha(1-\gamma^5)(\pslash_{\ell^-}+m_\ell)
\gamma^\beta(1-\gamma^5)v(p_{\bar{\nu}_\ell})\,,\label{eq:L2}
\end{eqnarray}
where $\vec{p}_{\ell^+}=\vec{p}_K-\vec{p}_{\nu_\ell}-\vec{p}_{n}$ in Eq.(\ref{eq:L1}) and $\vec{p}_{\ell^-}=\vec{p}_K-\vec{p}_{\bar{\nu}_\ell}-\vec{p}_{n_s}$ in Eq.(\ref{eq:L2}). In both cases the energy of the charged lepton is given by $E_{\ell^\pm}=\sqrt{\vec{p}_{\ell^\pm}^{\hspace{2.5pt}2}+m_\ell^2}$.

In deriving Eq.\,(\ref{eq:amplitudeM}) we have used the space-time translation property
\begin{equation}
\langle f\,|\,O(t,\vec{x})\,|\,i\rangle=e^{-i(E_i-E_f)t+i(\vec{p}_i-\vec{p}_f)\cdot\vec{x}}
\langle f\,|\,O(0,\vec{0})\,|\,i\rangle\label{eq:translation}\,,
\end{equation}
where $(E_{i},\vec{p}_{i})$ and $(E_{f},\vec{p}_{f})$ are the four-momenta of the initial and final states respectively, and have defined 
the quark $V-A$ currents by 
\be
O_{s,\alpha}^{(q)}=\bar{s}\gamma_\alpha(1-\gamma_5)q,\quad
O_{d,\beta}^{(q)}=\bar{q}\gamma_\beta(1-\gamma_5)d.
\ee

The principal objective of this paper is to demonstrate how to obtain the expression on the right-hand side of Eq.\,(\ref{eq:amplitudeM}) from computations of correlation functions on a finite Euclidean lattice with only exponentially small finite-volume effects. We explain how to achieve this in the following section.

\begin{center}
\begin{figure}[t!]
\includegraphics[width=0.43\hsize]{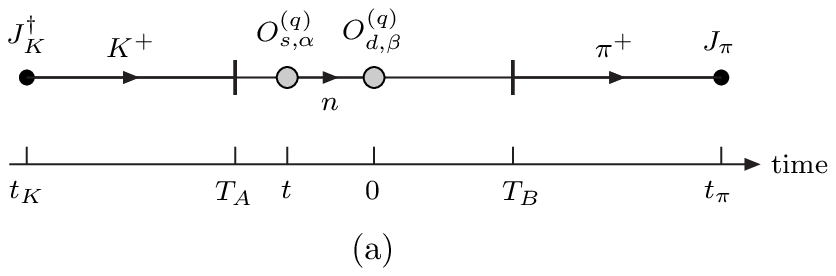}\hspace{0.5in}
\includegraphics[width=0.43\hsize]{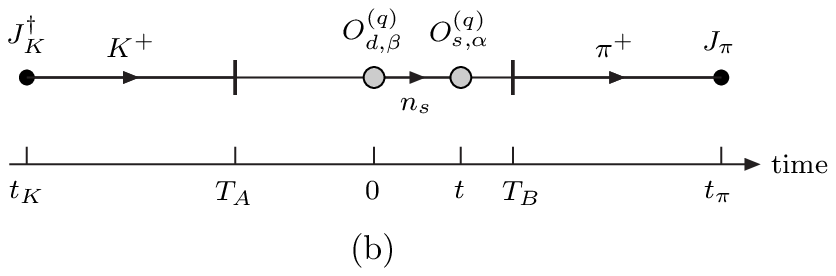}
\caption{Schematic drawing of the two time orderings in the correlation function. In (a) we have $t<0$ and in (b) we have $t>0$.}\label{fig:correlators}
\end{figure}
\end{center}

\section {$\mathbf{A^M_{q,\ell}}$ from Euclidean correlation functions}\label{sec:Euclidean}

Hadronic matrix elements are obtained in lattice QCD computations from calculations of finite-volume Euclidean correlation functions. In this section we present a detailed discussion of the evaluation of $A^M_{u,\ell}$, since the evaluation of $A_{c,\ell}^M$ is considerably more straightforward and can readily be deduced from that of $A^M_{u,\ell}$ (as we briefly explain at the appropriate points in the discussion). We consider separately the two time-orderings $t<0$ and $t>0$ corresponding to each of the two terms on the right-hand side of Eq.\,(\ref{eq:amplitudeM}). The correlation functions for the two time-orderings are sketched schematically in Fig.\,\ref{fig:correlators}.

\subsection{The time-ordering $\mathbf{t<0}$}
Consider the finite-volume Euclidean correlation function
\begin{eqnarray}
C_{\alpha\beta}^{(u) }(t,\vec{x})&=&
\sum_{\vec{x}_\pi,\vec{x}_K}\langle 0 |J_\pi(t_\pi,\vec{x}_\pi)\,
\,O_{d,\beta}^{(u) }(0,\vec{0}\,)\,O_{s,\alpha}^{(u) }(t,\vec{x})\,\,J_K^\dagger(t_K,\vec{x}_K)|0\rangle
e^{i\vec{p}_K\cdot \vec{x}_K}e^{-i\vec{p}_\pi\cdot\vec{x}_\pi}
\,,\label{eq:Cplus}
\end{eqnarray}
where $J_K^\dagger$ and $J_\pi$ are interpolating operators for the creation of a kaon and annihilation of a pion respectively. The correlation function in Eq.\,(\ref{eq:Cplus}) describes the creation of a kaon at a large negative time $t_K$, the insertion of the weak operators  $O_{s,\alpha}^{(u) }$ and $O_{d,\beta}^{(u) }$ at times $t$ and $0$ respectively, with $t<0$ and the annihilation of the pion at time $t_\pi\gg 0$. This is illustrated in Fig.\,\ref{fig:correlators}(a). For compactness of notation we suppress the dependence of $C_{\alpha\beta}^{(u) }(t)$ on $t_K,\,t_\pi$ and the momenta. In this section we show how to obtain the $A_{u,\ell}^{M,-}$ component of the amplitude from the evaluation of $C_{\alpha\beta}^{(u) }(t,\vec{x}\hspace{1pt})$ up to exponentially small finite-volume corrections.

Assuming, as is standard, that $t-t_k$ and $t_\pi$ are sufficiently large for the correlation function to be dominated by a kaon of momentum $\vec{p}_K$ propagating in the time interval $(t_K,t)$ and for a single pion to be propagating in the interval $(0,t_\pi)$ we have
\begin{eqnarray}
C_{\alpha\beta}^{(u) }(t,\vec{x})=Z_K Z_\pi \frac{e^{E_K t_K}}{2E_K}\frac{e^{-E_\pi t_\pi}}{2E_\pi}
\langle \pi(\vec{p}_\pi) |\,
O_{d,\beta}^{(u) }(0,\vec{0}\,)\,O_{s,\alpha}^{(u) }(t,\vec{x})\,|K(\vec{p}_K)\rangle
\,.\label{eq:Cplus2}
\end{eqnarray}
The energies of the kaon and pion, $E_K$ and $E_\pi$ respectively, and the matrix elements $Z_K=\langle K(\vec{p}_K)|J_K^\dagger(0)|0\rangle$ and $Z_\pi=\langle 0|J_\pi(0)|\pi(\vec p_\pi)\rangle$ can be obtained in the standard way from two-point meson correlation functions using our normalization conventions, { \it e.g.} for the
finite-volume state $|\pi(\vec p_\pi)\rangle$,
\begin{equation}
\langle \pi(\vec p_\pi^{\hspace{2.5pt}\prime})|\pi(\vec p_\pi)\rangle = 2E_\pi \left(\frac{L}{2\pi}\right)^{\!\!3}\delta_{\vec p_\pi^{\hspace{2.5pt}\prime} ,\vec p_\pi}.
\end{equation}
We then rewrite
Eq.\,(\ref{eq:Cplus2}) as
\begin{equation}
C_{\alpha\beta}^{(u) }(t,\vec{x})\equiv Z_{K\pi} H^{E,(u) }_{\alpha\beta}(t,\vec{x})\,,
\end{equation}
where
\begin{equation}
Z_{K\pi}=Z_K Z_\pi \frac{e^{E_K t_K}}{2E_K}\frac{e^{-E_\pi t_\pi}}{2E_\pi}\label{eq:ZKpi}
\end{equation}
and $H^{E,(u) }_{\alpha\beta}(t,\vec{x})$ is the Euclidean equivalent of the bilocal hadronic matrix element in Eq.\,(\ref{eq:HM}) at $t<0$
\begin{eqnarray}
H^{E,(u) }_{\alpha\beta}(t,\vec{x})&=&\langle\pi^+(p_\pi)|\big[\bar{u}(0)\gamma_\beta(1-\gamma_5)d(0)
\,\bar{s}(x)\gamma_\alpha
   (1-\gamma_5)u(x)\big]\,|K^+(p_K)\rangle
   \nonumber\\
   &&\hspace{-0.7in}=\sum_{n}\,\langle \pi(p_\pi)|O_{d,\beta}^{(u) }(0)|n(p_{n})\rangle\,\langle
n(p_{n})|O_{s,\alpha}^{(u) }(0)|K(p_K)\rangle\,
e^{i(\vec{p}_{K}-\vec{p}_n)\cdot\vec{x}}e^{-(E_{K}-E_n)t}\,,\label{eq:HEminus}
\end{eqnarray}
where the sum is over a complete set of non-strange states $|n\rangle$. The basis of the infinite-volume reconstruction method is that we perform the integral in Eq.\,(\ref{eq:H_and_L}) using the hadronic matrix element $H^{E,(u) }_{\alpha\beta}(t,\vec{x})$ calculated using lattice methods on a finite spatial
volume and the leptonic tensor $L^{E,\alpha\beta}$
calculated in an infinite spatial volume which for $t<0$ is given by
\begin{equation}\label{eq:LEminus}
L^{E,\alpha\beta}(x)=\int\frac{\dthree p_{\ell^+}}{(2\pi)^3}\,\frac{e^{(E_{\nu_\ell}+E_{\ell^+})t}\,
\,e^{-i(\vec{p}_{\ell^+}+\vec{p}_{\nu_\ell})\cdot\vec{x}}}{2E_{\ell^+}}\,
\bar{u}(p_{\nu_\ell})\gamma^\alpha(1-\gamma^5)(\pslash_{\ell^+}-m_\ell)
\gamma^\beta(1-\gamma^5)v(p_{\bar{\nu}_\ell})\,.
\end{equation}
In order to allow the external kaon and pion to propagate over sufficiently large time intervals to eliminate excited external states and obtain  $H^{E,(u) }_{\alpha\beta}(t,\vec{x})$ we imagine performing the time integration over the interval $(T_A,T_B)$, where $t_K\ll T_A\ll 0\ll T_B\ll t_\pi$,
as illustrated in Fig.\,\ref{fig:correlators}. In this subsection we are considering the contribution from the region $t<0$ and so the range of integration is $(T_A,0)$ and the aim here is to compute
\begin{equation}\label{eq:convolutionminus}
A^{E,-}_{u,\ell}=\int_{T_A}^0 dt \int_{L^3}\hspace{-3pt}\dthree x~H_{\alpha\beta}^{E,(u) }(t,\vec{x})L^{E,\alpha\beta}(t,\vec{x})\,,
\end{equation}
in such a way as to reproduce the first term on the right-hand side of Eq.\,(\ref{eq:amplitudeM}), $A^{M,-}_{u,\ell}$, with only exponentially small finite-volume effects\,\footnote{The $L^3$ suffix in Eq.\,(\ref{eq:convolutionminus}) indicates that the integral is performed over the finite spatial volume.}.
However, the presence of states $|n\rangle$ in the sum in the second line of Eq.\,(\ref{eq:HEminus}) with  energies which are smaller than those of the external states leads to exponentially growing terms in $|T_A|$ and power-law finite-volume effects. We therefore cannot simply evaluate the integral in Eq.\,(\ref{eq:convolutionminus}) using $H^{E,(u) }_{\alpha\beta}(t,\vec{x})$ computed directly on a finite Euclidean lattice for all $t\in (T_A,0)$ and a modified procedure must be introduced. We now explain in some detail the presence of power-law finite-volume effects and the exponentially growing behaviour with $|T_A|$ in Eq.\,(\ref{eq:convolutionminus}) together with our proposed method for eliminating them.

Using Eqs.\,(\ref{eq:HEminus}) and (\ref{eq:LEminus}) we see that the integration over time (under the sum over $|n\rangle$ and integration over $\vec{p}_{\ell^+}$) is given by
\begin{equation}\label{eq:tintminus}
\int_{T_A}^0dt~e^{(E_{n}+E_{\ell^+}+E_{\nu_\ell}-E_K)t}=\frac1{E_n+E_{\ell^+}+E_{\nu_\ell}-E_K}
\left[1-e^{-(E_{n}+E_{\ell^+}+E_{\nu_\ell}-E_K)|T_A|}\right].
\end{equation}
The difficulty arises because there are states $|n\rangle$ for which $E_{n}+E_{\ell^+}+E_{\nu_\ell}-E_K<0$ leading to an unphysical contribution that is exponentially growing in $T_A$ and power-law finite-volume effects due to the singularity in the denominator of the right-hand side of Eq.\,(\ref{eq:tintminus}). This singularity is present in the range of the summation over intermediate states. This demonstrates that it is not possible to achieve the stated aim by performing the  integral in Eq.\,(\ref{eq:convolutionminus}) using
$H_{\alpha\beta}^{E,(u) }(x)$, computed on the lattice over the full time interval $(T_A,0)$, and the infinite-volume leptonic tensor $L^{E,\alpha\beta}(t,\vec{x})$ in Eq.\,(\ref{eq:LEminus}). We now discuss the modifications we propose in order to complete the determination of the first term of the physical amplitude $A_{u,\ell}^{M,-}$, i.e. the first term on the right-hand side of Eq.\,(\ref{eq:amplitudeM}), with only exponential finite-volume effects.

We separate $H^{E,(u)}_{\alpha\beta}(t,\vec{x})$ into the contributions from the vacuum  and the hadronic intermediate states writing
\be\label{eq:separate}
H^{E,(u)}_{\alpha\beta}(t,\vec{x})=H_{\alpha\beta}^{\mathrm{vac}}(t,\vec{x})+H_{\alpha\beta}^{\mathrm{had},(u)}(t,\vec{x}),
\ee
where
\be\label{eq:Hvac}
H_{\alpha\beta}^{\mathrm{vac}}(t,\vec{x})\equiv\langle \pi|O_{d,\beta}^{(u)}(0)|0\rangle\langle 0|O_{s,\alpha}^{(u)}(0)|K\rangle
e^{i\vec{p}_K\cdot\vec{x}}e^{-E_Kt}.
\ee
The vacuum contribution $H_{\alpha\beta}^{\mathrm{vac}}(t,\vec{x})$ can be determined with only exponential finite-volume errors using the matrix elements $\langle \pi|O_{d,\beta}^{(u)}(0)|0\rangle^\mathrm{latt}$ and $\langle 0|O_{s,\alpha}^{(u)}(0)|K\rangle^\mathrm{latt}$ determined in lattice computations of two-point correlation functions in the standard way. We can therefore obtain the contribution from the purely leptonic intermediate state ($|n\rangle=|0\rangle$) to the physical amplitude in Eq.\,(\ref{eq:amplitudeM})
\begin{equation}\label{eq:Avac}
A^{\mathrm{vac}}=\frac{\langle \pi|O_{d,\beta}^{(u)}(0)|0\rangle^\mathrm{latt}\langle 0|O_{s,\alpha}^{(u)}(0)|K\rangle^\mathrm{latt}}
{E_{\ell^+}+E_{\nu_\ell}-E_K-i\varepsilon}L^{\alpha\beta}_1(\vec{0}\,)\,.
\end{equation}
The subtraction of the vacuum contribution analogous to that needed to obtain
$H^{\mathrm{had},(u)}$ from Eq.\,(\ref{eq:separate}) has been performed
successfully in
a study of neutrinoless double $\beta$-decay\,\cite{Tuo:2019bue} and here we also envisage exploiting the statistical correlations between $H^{E,(u)}_{\alpha\beta}$ and $H_{\alpha\beta}^{\mathrm{vac}}$ in order to obtain sufficiently precise values of $H_{\alpha\beta}^{\mathrm{had},(u)}$.

We now consider $H_{\alpha\beta}^{\mathrm{had},(u)}(t,\vec{x})$ for $t<0$, obtained after subtracting the vacuum contribution from $H^{E,(u)}_{\alpha\beta}(t,\vec{x})$. We assume that for some sufficiently large $|t_s|$ the hadronic factor $H_{\alpha\beta}^{\mathrm{had},(u)}(t,\vec{x})$ with $|t|\ge|t_s|$ 
is dominated by the $|\pi^0\rangle$ intermediate state, so that in particular
\begin{eqnarray}
H^{\mathrm{had},(u)}_{\alpha\beta}(t_s,\vec{x})&\simeq&\int
\frac{\dthree p_{\pi^0}}{(2\pi)^3}\,\frac1{2E_{\pi^0}}~\langle \pi(p_\pi)|O_{d,\beta}^{(u) }(0)|\pi^0(p_{\pi^0})\rangle\,\langle
\pi^0(p_{\pi^0})|O_{s,\alpha}^{(u) }(0)|K(p_K)\rangle\,\times\nn\\ 
&&\hspace{1in}e^{i(\vec{p}_{K}-\vec{p}_{\pi^0})\cdot\vec{x}}e^{-(E_{K}-E_{\pi^0})t_s}\,,\label{eq:ground_dominance}
\end{eqnarray}
where the $\simeq$ symbol indicates the equality of the two sides of the equation up to excited-state contributions which are assumed to be negligible. Although $|t_s|$ is large enough for the ground state to dominate for $|t|\ge |t_s|$, it is nevertheless finite so that the finite-volume corrections in $H^{\mathrm{had},(u)}_{\alpha\beta}(t_s,\vec{x})$ are exponentially suppressed and we have therefore replaced the sum over finite-volume $|\pi^0\rangle$ by the infinite-volume phase-space integral.

In the original presentation of the IVR method\,\cite{Feng:2018qpx}, the integration region over $t$, i.e. $t\in(-\infty,0)$, is divided into two intervals $(t_s,0)$ and $(-\infty,t_s)$, labeled as regions $s$ (for \emph{short}) and $l$ (\emph{long}) respectively. Here we start by evaluating $\mathcal{I}^{(s)}$, the integral over the region $t\in(t_s,0)$ (see Eq.\,(\ref{eq:Isdef}) below). We then show that the corresponding contribution to the physical amplitude $A_{u,\ell}^{M,-}$ (see Eqs.\,(\ref{eq:amplitudeM}) and (\ref{eq:amplitudeMmp})) is obtained from $\mathcal{I}^{(s)}+\tilde{\mathcal{I}}^{(l)}$, where $\tilde{\mathcal{I}}^{(l)}$ is an appropriately modified contribution from the integration region $(-\infty,t_s)$. The hadronic components of $\mathcal{I}^{(s)}$ and $\tilde{\mathcal{I}}^{(l)}$ can both be determined from lattice computations.

The integration over $t$ in the interval $(t_s,0)$ is defined by
\begin{eqnarray}
\mathcal{I}^{(s)}&\equiv&\int_{t_s}^0dt\int_{L^3}
\dthree x~H_{\alpha\beta}^{\mathrm{had},(u)}(t,\vec{x})L^{E,\alpha\beta}(t,\vec{x})\nn\\ 
&\simeq&\int_{t_s}^0dt\int_{\infty}
\dthree x~H_{\alpha\beta}^{\mathrm{had},(u)}(t,\vec{x})L^{E,\alpha\beta}(t,\vec{x})\,,\label{eq:Isdef}
\end{eqnarray}
since for finite $t_s$ the finite-volume effects are exponentially small\,\footnote{The suffix $\infty$ in Eq.\,(\ref{eq:Isdef}) indicates that the integral is performed in infinite volume.}.
Nevertheless $\mathcal{I}^{(s)}$ does not reproduce the corresponding contribution to the Minkowski amplitude, $A_{u,\ell}^{(M,-)}$ defined in Eq.\,(\ref{eq:amplitudeMmp}). Instead $\mathcal{I}^{(s)}$ is given by 
\begin{equation}
\mathcal{I}^{(s)}
=\int d\phi_n\,\frac{\langle \pi|O_{d,\beta}^{(u) }(0)|n(\vec{p}_n)\rangle\langle
n(\vec{p}_n)|O_{s,\alpha}^{(u) }(0)|K\rangle}{E_n+E_{\ell^+}+E_{\nu_\ell}-E_K}\,L_1^{\alpha\beta}(\vec{p}_n)\left[1-e^{-(E_{n}+E_{\ell^+}+E_{\nu_\ell}-E_K)|t_s|}\right]\label{eq:Is}\,,
\end{equation}
where a sum over intermediate states $|n\rangle$ is implied.
For excited states with $E_n+E_{\ell^+}+E_{\nu_\ell}-E_K>0$ the contributions
from Euclidean matrix elements reproduce the corresponding Minkowski results up to the
exponentially suppressed term $e^{-(E_{n}+E_{\ell^+}+E_{\nu_\ell}-E_K)|t_s|}$. This is not the case however, for the ground-state, $|\pi^0\rangle$, contribution which has to be treated differently.
(Note that even in this case, the integrand on the right-hand side of Eq.\,(\ref{eq:Is}) has no singularity at $E_{\pi^0}+E_{\ell^+}+E_{\nu_\ell}-E_K=0$ since the numerator also vanishes at this point.) 
 
In order to reproduce $A^{M,-}_{u,\ell}$ we must remove the $|\pi^0\rangle$ contribution from $\mathcal{I}^{(s)}$ in (\ref{eq:Is}), and replace it by the corresponding term (i.e. the term with $|n\rangle=|\pi^0\rangle$) in Eq.\,(\ref{eq:amplitudeM}). To this end we define the quantity
$\tilde{\mathcal{I}}^{(l)}$ by \footnote{The  tilde on $\tilde{\mathcal{I}}^{(l)}$ is introduced to denote the fact that $\tilde{\mathcal{I}}^{(l)}$ is not simply the integral over the region $t<t_s$.}
\ba
\tilde{\mathcal{I}}^{(l)}&=&\int\frac{\dthree p_{\pi^0}}{(2\pi)^3}\,\frac{1}{2E_{\pi^0}}
\langle\pi|O_{d,\beta}^{(u) }(0)|\pi^0(\vec{p}_{\pi^0}\vs)\rangle\langle
\pi^0(\vec{p}_{\pi^0}\vs)|O_{s,\alpha}^{(u) }(0)|K\rangle
L_1^{\alpha\beta}(\vec{p}_{\pi^0}\vs)\times
\nn\\
&&
\hspace{1cm}
\left[\frac{1}{E_{\pi^0}+E_{\ell^+}+E_{\nu_\ell}-E_K-i\varepsilon}-\frac{1-e^{-(E_{\pi^0}+E_{\ell^+}+E_{\nu_\ell}-E_K)|t_s|}}{E_{\pi^0}+E_{\ell^+}+E_{\nu_\ell}-E_K-i\varepsilon}\right]\,,
\label{eq:Ilmodify}\\
\rule[0in]{0in}{0.4in}&=&\int\frac{\dthree p_{\pi^0}}{(2\pi)^3}\,\frac{1}{2E_{\pi^0}}\langle\pi|O_{d,\beta}^{(u) }(0)|\pi^0(\vec{p}_{\pi^0}\vs)\rangle\langle
\pi^0(\vec{p}_{\pi^0}\vs)|O_{s,\alpha}^{(u) }(0)|K\rangle
L_1^{\alpha\beta}(\vec{p}_{\pi^0}\vs)\times\nn\\
&&\hspace{2cm}\frac{e^{-(E_{\pi^0}+E_{\ell^+}+E_{\nu_\ell}-E_K)|t_s|}}{E_{\pi^0}+E_{\ell^+}+E_{\nu_\ell}-E_K-i\varepsilon}\,,\label{eq:finaltildeIl}
\ea
where $E_{\pi^0}=\sqrt{m_\pi^2+\vec{p}_{\pi^0}^{\,\,2}}$.
In the second line of Eq.\,(\ref{eq:Ilmodify})
the first term is the Minkowski contribution from the $|\pi^0\rangle$ intermediate state (see Eq.\,(\ref{eq:amplitudeM}))
and the second term is the $|\pi^0\rangle$ contribution to 
$\mathcal{I}^{(s)}$ (see Eq.\,(\ref{eq:Is})). Since in this second term there is no singularity, it is possible and also convenient to add $-i\varepsilon$ to the denominator. 
In this way we remove the unphysical contribution  in $\mathcal{I}^{(s)}$ and replace it with the missing term in the physical amplitude.
Combining Eqs.\,(\ref{eq:ground_dominance}) and (\ref{eq:finaltildeIl})
we have
\be\label{eq:Iltilde}
\tilde{\mathcal{I}}^{(l)}=\int
\dthree x\,H_{\alpha\beta}^{\mathrm{had},(u)}(t_s,\vec{x})\tilde{L}_1^{\alpha\beta}(t_s,\vec{x}),
\ee
where $\tilde{L}_1^{\alpha\beta}(t_s,\vec{x})$ is defined as
\be\label{eq:L1tilde}
\tilde{L}_1^{\alpha\beta}(t_s,\vec{x})=\int\frac{\dthree p_{\ell^+}}{(2\pi)^3}~e^{-i(\vec{p}_{\ell^+}+\vec{p}_{\nu_\ell})\cdot\vec{x}}\,L_1^{\alpha\beta}(\vec{p}_{\pi^0}\!)\,\frac{e^{-(E_{\ell^+}+E_{\nu_\ell})|t_s|}}{E_{\pi^0}+E_{\ell^+}+E_{\nu_\ell}-E_K-i\epsilon}\,.
\ee
In the integrand of Eq.\,(\ref{eq:L1tilde}), $\vec{p}_{\pi_0}=\vec{p}_K-\vec{p}_{\ell^+}-\vec{p}_{\nu_\ell}$ and $E_{\pi_0}=\sqrt{\vec{p}_{\pi_0}^{\hspace{2.5pt}2}+m_{\pi_0}^2}$. Thus we see that the quantities $\mathcal{I}^{(s)}$ and $\tilde{\mathcal{I}}^{(l)}$ can be
approximated using the quantities
$H_{\alpha\beta}^{\mathrm{had}}(t,\vec{x})$ calculated for $|t|\le |t_s|$ in a lattice computation as inputs. The finite-volume effects induced by this approximation are exponentially suppressed.

Combining the contribution from the vacuum intermediate state in Eq.(\ref{eq:Avac}), with those from the hadronic intermediate states in Eqs.\,(\ref{eq:Isdef}) and (\ref{eq:Iltilde}), we obtain the contribution to the physical amplitude in Eq.(\ref{eq:amplitudeM}) from the region $x_0<0$ with only exponential finite-volume corrections:
\begin{equation}
A^{M,-}_{u,\ell}=A^{\mathrm{vac}}+{\cal I}^{(s)}+\tilde{\mathcal{I}}^{(l)}\,.
\end{equation}

The contribution $A^{M,-}_{c,\ell}$ is much more straightforward to evaluate as the intermediate states now have charm quantum number $C=1$, and so have larger energies than $m_K$. 
In this case one simply performs the integral 
\begin{equation}
A^{E,-}_{c,\ell}=\int_{T_A}^0 dt\int\dthree x\,H_{\alpha\beta}^{E,(c) }(t,\vec{x})L^{E,\alpha\beta}(t,\vec{x})
\end{equation}
with the hadronic matrix elements computed directly in lattice QCD. In this case $A^{E,-}_{c,\ell}=A^{M,-}_{c,\ell}$ up to exponential finite-volume effects.

\subsection{The time-ordering $\mathbf{t>0}$}\label{subsec:tplus}

We now consider the case $t>0$ and the evaluation of $A^{M,+}_{u,\ell}$,  
for which the elimination of the power-law finite-volume effects is a little more straightforward but which nevertheless contains a new subtlety. We start by following the same steps as for $t<0$, relating the Euclidean correlation function illustrated in Fig.\,\ref{fig:correlators}(b) to the bilocal hadronic matrix element:
\begin{eqnarray}
C_{\alpha\beta}^{(u)}(t,\vec{x})&=&
\sum_{\vec{x}_\pi,\vec{x}_K}\langle 0 |J_\pi(t_\pi,\vec{x}_\pi)\,
\,O_{s,\alpha}^{(u)}(t,\vec{x})\,O_{d,\beta}^{(u)}(0,\vec{0}\,)\,J_K^\dagger(t_K,\vec{x}_K)|0\rangle
e^{i\vec{p}_K\cdot \vec{x}_K}e^{-i\vec{p}_\pi\cdot\vec{x}_\pi}\nonumber\\
&=&Z_{K\pi} H^{E,(u) }_{\alpha\beta}(t,\vec{x})\,,
\label{eq:Cminus}
\end{eqnarray}
where $Z_{K\pi}$ is given in Eq.\,(\ref{eq:ZKpi}) and for 
$t>0$\,,
\begin{eqnarray}
H^{E,(u)}_{\alpha\beta}(t,\vec{x})&=&\langle\pi^+(p_\pi)|\big[\bar{s}(x)\gamma_\alpha
   (1-\gamma_5)u(x)\big]\,\big[\bar{u}(0)\gamma_\beta(1-\gamma_5)d(0)\big]\,|K^+(p_K)\rangle
   \nonumber\\
   &&\hspace{-0.7in}=\sum_{n_s}\,\langle \pi(p_\pi)|O_{s,\alpha}^{(u) }(0)|n_s(p_{n_s})\rangle\langle
n_s(p_{n_s})|O_{d,\beta}^{(u) }(0)|K(p_K)\rangle\,
e^{i(\vec{p}_{n_s}-\vec{p}_\pi)\cdot\vec{x}}e^{-(E_{n_s}-E_\pi)t}\,,
\end{eqnarray}
and the sum is over the finite-volume multi-hadron strangeness $S=1$ states $|n_s\rangle$. There are therefore no on-shell intermediate states $|n_s\rangle$ and consequently there is no unphysical exponentially growing behaviour in $T_B$ in the integral over $t$. Nevertheless there is a subtlety which must be taken into account. Consider the Type 1 diagram in Fig.\,\ref{fig:Wbox}. Although we have drawn the two loops as only being connected by a lepton propagator, it is to be understood that gluonic and vacuum polarization effects, although not drawn explicitly, are also implicitly included.   However, the functional integral over the gluon and sea-quark fields contains contributions 
in which the strong-interaction effects are restricted to each loop separately and do not connect the two loops. For these contributions there is no suppression in $|\vec{x}|$  when evaluating $H^{E,(u)}_{\alpha\beta}(t,\vec{x})$ and they need to be treated separately in an analogous way to the vacuum contribution in Eq.\,(\ref{eq:Hvac}). We call these contributions \emph{disconnected}\,\footnote{In lattice QCD literature ``disconnected" frequently refers to diagrams in which quark loops are only connected by gluons. We stress that our use of ``disconnected" here is different and denotes diagrams with no strong interactions between the quark loops.}.

Analogously to Eq.\,(\ref{eq:separate}) we write for $t>0$,
\be\label{eq:separate2}
H^{E,(u)}_{\alpha\beta}(t,\vec{x})=H_{\alpha\beta}^{\mathrm{disc}}(t,\vec{x})+H_{\alpha\beta}^{\mathrm{conn},(u)}(t,\vec{x})
\ee
where the labels {\footnotesize disc} and {\footnotesize conn} represent the disconnected and connected contributions respectively. The disconnected contribution has the same form as $H^{\mathrm{vac}}(t,\vec{x})$, but now $t>0$:
\be\label{eq:Hdisc}
H_{\alpha\beta}^{\mathrm{disc}}(t,\vec{x})\equiv\langle 0|O_{s,\alpha}^{(u)}(0)|K\rangle\,\langle \pi|O_{d,\beta}^{(u)}(0)|0\rangle
e^{i\vec{p}_K\cdot\vec{x}}e^{-E_Kt}.
\ee
Each of the two local matrix elements in Eq.\,(\ref{eq:Hdisc}) can be computed independently as an average over the gauge configurations with only exponential finite-volume corrections.
We write the corresponding contribution to the amplitude in the form
\begin{equation}\label{eq:A2disc}
A_{u,\ell}^{\mathrm{disc}}=\frac{\langle \pi|O_{d,\beta}^{(u)}(0)|0\rangle\langle 0|O_{s,\alpha}^{(u)}(0)|K\rangle}{(E_K+E_\pi)+E_{\bar{\nu}_\ell}
+E_{\ell^-}-E_K}\,L_2(\vec{p}_K+\vec{p}_\pi)
\end{equation}
to demonstrate that it is the disconnected contribution to $A^{M,+}_{u,\ell}$ in Eq.\,(\ref {eq:amplitudeM}). Note that for the disconnected contribution $p_{n_s}=p_K+p_\pi$.

For the connected contribution, there are two important points to note, both resulting from the observation that the intermediate states $|n_s\rangle$ all have energies which are larger than those of the external states. The first point is that the finite-volume effects in $H^{\mathrm{conn},(u)}_{\alpha\beta}(t,\vec{x})$ computed on a finite lattice are exponentially suppressed. The second related point is that even in infinite volume this hadronic matrix element is exponentially suppressed at large $t$ and $|\vec{x}|$.

In evaluating the contribution to the integral of Eq.\,(\ref{eq:H_and_L}) we need to combine $H^{\mathrm{conn},(u)}_{\alpha\beta}(t,\vec{x})$ with the corresponding leptonic tensor $L^{E,\alpha\beta}(t,\vec{x})$, where for $t>0$
\begin{equation}\label{eq:LEplus}
L^{E,\alpha\beta}(x)=\int\frac{\dthree p_{\ell^-}}{(2\pi)^3}\,\frac{e^{-(E_{\nu_\ell}-E_{\ell^-})t}\,
\,e^{i(\vec{p}_{\ell^-}-\vec{p}_{\nu_\ell})\cdot\vec{x}}}{2E_{\ell^-}}\,
\bar{u}(p_{\nu_\ell})\gamma^\alpha(1-\gamma^5)(\pslash_{\ell^-}+m_\ell)
\gamma^\beta(1-\gamma^5)v(p_{\bar{\nu}_\ell})\,.
\end{equation}
In this subsection we are considering the contribution from the region $t>0$, so the range of integration is $(0,T_B)$ and we arrive at the following contribution to the decay amplitude:
\begin{equation}\label{eq:amplitudeEplus}
A^\mathrm{conn}_{u,\ell}=\int_0^{T_B} dt \int_{L^3}\hspace{-5pt}\dthree x~H^{\mathrm{conn},(u)}_{\alpha\beta}(t,\vec{x})\,L^{E,\alpha\beta}(t,\vec{x})\,.
\end{equation}
The contribution $A^{\mathrm{conn}}_{u,\ell}$ is equal to the connected contribution to $A^{M,+}_{u,\ell}$ up to exponentially suppressed terms in the volume and in $T_B$.

Combining Eqs.\,(\ref{eq:A2disc}) and (\ref{eq:amplitudeEplus}) we obtain
\begin{equation}
A^{M,+}_{u,\ell}=A^\mathrm{disc}_{u,\ell}+A^\mathrm{conn}_{u,\ell}\,,
\end{equation}
where the equality holds up to exponentially small finite-volume corrections.

The evaluation of the corresponding contribution from the charmed intermediate states, i.e. to $A^{M^+}_{c,\ell}$, follows in the same way except that there are no Type 1 diagrams and hence there is no disconnected contribution.

The discussion of the contribution from the region $t>0$ is considerably simplified because the necessary presence of multi-hadron $S=1$ intermediate states implies that there are no power-law finite-volume effects arising from on-shell intermediate states.
In addition, the use of the infinite-volume leptonic tensor (\ref{eq:LEplus}) in the integration in Eq.\,(\ref{eq:amplitudeEplus}) avoids power-law finite-volume effects which would arise due to the factor of $1/2E_{l^-}$ in the difference between a finite-volume sum over $\vec{p}_{\ell^-}$ and the corresponding infinite-volume integration. 
The above discussion is a particular illustration of how to avoid power-law finite-volume corrections in the second term on the right-hand side of Eq.\,(\ref{eq:twosingularities}), which applies to general processes with a massless (or almost massless) leptonic propagator. 

\subsection{Summary}
In summary therefore, we propose to calculate the decay amplitude
$A^{M}_{u,\ell}$, using the following form where all the hadronic quantities can be obtained from lattice simulations with exponential finite-volume effects:
\ba
A^M_{u,\ell}&=&
\frac{\langle \pi|O_{d,\beta}^{(u)}(0)|0\rangle^{\mathrm{latt}}\langle 0|O_{s,\alpha}^{(u)}(0)|K\rangle^{\mathrm{latt}}}
{E_{\ell^+}+E_{\nu_\ell}-E_K-i\varepsilon}\,L_1^{\alpha\beta}(\vec{0}\hspace{1.5pt})
+\int_{t_s}^0 dt \int_{L^3}\hspace{-0.1in} \dthree x\,H_{\alpha\beta}^{\mathrm{had,(u),latt}}(t,\vec{x})L^{E,\alpha\beta}(t,\vec{x})
\nn\\
&&\hspace{-0.5in}+\int_{L^3}\hspace{-0.1in}
\dthree x~H_{\alpha\beta}^{\mathrm{had,(u),latt}}(t_s,\vec{x})\tilde{L}_1^{\alpha\beta}(t_s,\vec{x})
+\,\frac{\langle \pi|O_{d,\beta}^{(u)}(0)|0\rangle^{\mathrm{latt}}\langle 0|O_{s,\alpha}^{(u)}(0)|K\rangle^{\mathrm{latt}}}{(E_K+E_\pi)+E_{\bar{\nu}_\ell}
+E_{\ell^-}-E_K}\,L^{\alpha\beta}_2(\vec{p}_K+\vec{p}_\pi)\nn\\ 
&&\hspace{0.4in}+\int_{0}^{T_B}
dt\int_{L^3}\hspace{-0.1in} \dthree x~H_{\alpha\beta}^{\mathrm{conn,(u),latt}}(t,\vec{x})L^{E,\alpha\beta}(t,\vec{x}\hspace{1pt})\,.\label{eq:master_formula}
\ea 
The first three terms on the right-hand side of Eq.\,(\ref{eq:master_formula}) come from the region $t<0$ and are respectively (i) the vacuum contribution obtained using $H^{\mathrm{vac}}_{\alpha\beta}$ in Eq.\,(\ref{eq:Hvac}) and the contributions from (ii) $\mathcal{I}_s$ in Eq.\,(\ref{eq:separate}) and (iii) $\tilde{\mathcal{I}}_l$ in Eqs.\,(\ref{eq:Iltilde}) and (\ref{eq:L1tilde}). The final two terms in Eq.\,(\ref{eq:master_formula}) come from the region 
region $t>0$ and are the (iv) disconnected contributions, see Eq.\,(\ref{eq:A2disc}) and (v) the connected contributions in Eq.\,(\ref{eq:amplitudeEplus}). The superscript {\footnotesize $\mathrm{latt}$} underlines the observation that the local and bilocal matrix elements are calculable in a lattice computation with only exponential finite-volume effects.

The focus of this paper has been on the demonstration of the presence of power-law finite-volume effects in the amplitudes $A^M_{u,\ell}$ for rare-kaon decays $K\to\pi\nu_\ell\bar\nu_\ell$ and the presentation of a proposed method to eliminate them. Such effects are absent from $A^M_{c,\ell}$, where the intermediate states carry $C=1$ charm quantum number and Type 1 diagrams do not contribute. There are therefore no vacuum or disconnected contributions, nor any from other intermediate states lighter then the kaon and so $H_{\alpha\beta}^{E,(c) }(t,\vec{x})$ as computed on a finite Euclidean lattice can be used directly in the integral $\int_{T_A}^{T_B}dt \int_{L^3}\dthree x\,H^{E,(c)}_{\alpha\beta}(t,\vec{x})\,L^{E,\alpha\beta}(t,\vec{x})$ to obtain $A^M_{c,\ell}$.


An important point to note is that it still requires further investigations to extend the method developed in Ref.\,\cite{Feng:2018qpx} and in this work to multi-hadron intermediate states with energies smaller than the external ones, as such states induce branch cuts which cannot be simply described by discrete QCD eigenstates in infinite volume. 

\section{Conclusion}\label{sec:concs}

In this work, we extend the \emph{infinite-volume reconstruction} method proposed in Ref.\,\cite{Feng:2018qpx} to long-distance processes with massless (or almost massless) leptonic propagators. Using the rare 
 $K^+\to\pi^+\nu_\ell\bar{\nu}_\ell$ decay as an example, we show that the power-law finite-volume effects induced by the massless electron can safely be removed using the form in Eq.\,(\ref{eq:master_formula}) in lattice computations. 
A similar approach has been applied to the amplitude for neutrinoless
double-$\beta$ decay in which there is the propagator of a massless
neutrino\,\cite{Tuo:2019bue}. We are also performing exploratory studies with
the aim of extending the method to the evaluation of  electromagnetic
corrections to leptonic and semileptonic decays~\cite{Feng:2020mmb} and applying it in numerical lattice QCD calculations.

\begin{acknowledgments}
We gratefully acknowledge many helpful discussions with our colleagues from the
RBC-UKQCD collaboration. N.H.C. was supported in part by U.S. DOE grant DE-SC0011941, X.F. 
by the NSFC of China under Grant No. 11775002, L.C.J. by DOE grant
DE-SC0010339 and DE-SC0021147 and C.T.S. by STFC (UK) grant\,ST/P000711/1 and by an Emeritus Fellowship from the Leverhulme Trust.
\end{acknowledgments}

\bibliography{paper}

\begin{thebibliography}{56}%
\makeatletter
\providecommand \@ifxundefined [1]{%
 \@ifx{#1\undefined}
}%
\providecommand \@ifnum [1]{%
 \ifnum #1\expandafter \@firstoftwo
 \else \expandafter \@secondoftwo
 \fi
}%
\providecommand \@ifx [1]{%
 \ifx #1\expandafter \@firstoftwo
 \else \expandafter \@secondoftwo
 \fi
}%
\providecommand \natexlab [1]{#1}%
\providecommand \enquote  [1]{``#1''}%
\providecommand \bibnamefont  [1]{#1}%
\providecommand \bibfnamefont [1]{#1}%
\providecommand \citenamefont [1]{#1}%
\providecommand \href@noop [0]{\@secondoftwo}%
\providecommand \href [0]{\begingroup \@sanitize@url \@href}%
\providecommand \@href[1]{\@@startlink{#1}\@@href}%
\providecommand \@@href[1]{\endgroup#1\@@endlink}%
\providecommand \@sanitize@url [0]{\catcode `\\12\catcode `\$12\catcode
  `\&12\catcode `\#12\catcode `\^12\catcode `\_12\catcode `\%12\relax}%
\providecommand \@@startlink[1]{}%
\providecommand \@@endlink[0]{}%
\providecommand \url  [0]{\begingroup\@sanitize@url \@url }%
\providecommand \@url [1]{\endgroup\@href {#1}{\urlprefix }}%
\providecommand \urlprefix  [0]{URL }%
\providecommand \Eprint [0]{\href }%
\providecommand \doibase [0]{http://dx.doi.org/}%
\providecommand \selectlanguage [0]{\@gobble}%
\providecommand \bibinfo  [0]{\@secondoftwo}%
\providecommand \bibfield  [0]{\@secondoftwo}%
\providecommand \translation [1]{[#1]}%
\providecommand \BibitemOpen [0]{}%
\providecommand \bibitemStop [0]{}%
\providecommand \bibitemNoStop [0]{.\EOS\space}%
\providecommand \EOS [0]{\spacefactor3000\relax}%
\providecommand \BibitemShut  [1]{\csname bibitem#1\endcsname}%
\let\auto@bib@innerbib\@empty
\bibitem [{\citenamefont {Feng}\ and\ \citenamefont
  {Jin}(2019)}]{Feng:2018qpx}%
  \BibitemOpen
  \bibfield  {author} {\bibinfo {author} {\bibfnamefont {X.}~\bibnamefont
  {Feng}}\ and\ \bibinfo {author} {\bibfnamefont {L.}~\bibnamefont {Jin}},\
  }\href {\doibase 10.1103/PhysRevD.100.094509} {\bibfield  {journal} {\bibinfo
   {journal} {Phys. Rev. D}\ }\textbf {\bibinfo {volume} {100}},\ \bibinfo
  {pages} {094509} (\bibinfo {year} {2019})},\ \Eprint
  {http://arxiv.org/abs/1812.09817} {arXiv:1812.09817 [hep-lat]} \BibitemShut
  {NoStop}%
\bibitem [{\citenamefont {Aoki}\ \emph {et~al.}(2020)\citenamefont {Aoki} \emph
  {et~al.}}]{Aoki:2019cca}%
  \BibitemOpen
  \bibfield  {author} {\bibinfo {author} {\bibfnamefont {S.}~\bibnamefont
  {Aoki}} \emph {et~al.} (\bibinfo {collaboration} {Flavour Lattice Averaging
  Group}),\ }\href {\doibase 10.1140/epjc/s10052-019-7354-7} {\bibfield
  {journal} {\bibinfo  {journal} {Eur. Phys. J. C}\ }\textbf {\bibinfo {volume}
  {80}},\ \bibinfo {pages} {113} (\bibinfo {year} {2020})},\ \Eprint
  {http://arxiv.org/abs/1902.08191} {arXiv:1902.08191 [hep-lat]} \BibitemShut
  {NoStop}%
\bibitem [{\citenamefont {Christ}\ \emph {et~al.}(2013)\citenamefont {Christ},
  \citenamefont {Izubuchi}, \citenamefont {Sachrajda}, \citenamefont {Soni},\
  and\ \citenamefont {Yu}}]{Christ:2012se}%
  \BibitemOpen
  \bibfield  {author} {\bibinfo {author} {\bibfnamefont {N.~H.}\ \bibnamefont
  {Christ}}, \bibinfo {author} {\bibfnamefont {T.}~\bibnamefont {Izubuchi}},
  \bibinfo {author} {\bibfnamefont {C.~T.}\ \bibnamefont {Sachrajda}}, \bibinfo
  {author} {\bibfnamefont {A.}~\bibnamefont {Soni}}, \ and\ \bibinfo {author}
  {\bibfnamefont {J.}~\bibnamefont {Yu}} (\bibinfo {collaboration} {RBC,
  UKQCD}),\ }\href {\doibase 10.1103/PhysRevD.88.014508} {\bibfield  {journal}
  {\bibinfo  {journal} {Phys. Rev.}\ }\textbf {\bibinfo {volume} {D88}},\
  \bibinfo {pages} {014508} (\bibinfo {year} {2013})},\ \Eprint
  {http://arxiv.org/abs/1212.5931} {arXiv:1212.5931 [hep-lat]} \BibitemShut
  {NoStop}%
\bibitem [{\citenamefont {Bai}\ \emph {et~al.}(2014)\citenamefont {Bai},
  \citenamefont {Christ}, \citenamefont {Izubuchi}, \citenamefont {Sachrajda},
  \citenamefont {Soni},\ and\ \citenamefont {Yu}}]{Bai:2014cva}%
  \BibitemOpen
  \bibfield  {author} {\bibinfo {author} {\bibfnamefont {Z.}~\bibnamefont
  {Bai}}, \bibinfo {author} {\bibfnamefont {N.~H.}\ \bibnamefont {Christ}},
  \bibinfo {author} {\bibfnamefont {T.}~\bibnamefont {Izubuchi}}, \bibinfo
  {author} {\bibfnamefont {C.~T.}\ \bibnamefont {Sachrajda}}, \bibinfo {author}
  {\bibfnamefont {A.}~\bibnamefont {Soni}}, \ and\ \bibinfo {author}
  {\bibfnamefont {J.}~\bibnamefont {Yu}},\ }\href {\doibase
  10.1103/PhysRevLett.113.112003} {\bibfield  {journal} {\bibinfo  {journal}
  {Phys. Rev. Lett.}\ }\textbf {\bibinfo {volume} {113}},\ \bibinfo {pages}
  {112003} (\bibinfo {year} {2014})},\ \Eprint {http://arxiv.org/abs/1406.0916}
  {arXiv:1406.0916 [hep-lat]} \BibitemShut {NoStop}%
\bibitem [{\citenamefont {Christ}\ and\ \citenamefont
  {Bai}(2016)}]{Christ:2015phf}%
  \BibitemOpen
  \bibfield  {author} {\bibinfo {author} {\bibfnamefont {N.~H.}\ \bibnamefont
  {Christ}}\ and\ \bibinfo {author} {\bibfnamefont {Z.}~\bibnamefont {Bai}},\
  }\bibfield  {booktitle} {\emph {\bibinfo {booktitle} {{Proceedings, 33rd
  International Symposium on Lattice Field Theory (Lattice 2015): Kobe, Japan,
  July 14-18, 2015}}},\ }\href {\doibase 10.22323/1.251.0342} {\bibfield
  {journal} {\bibinfo  {journal} {PoS}\ }\textbf {\bibinfo {volume}
  {LATTICE2015}},\ \bibinfo {pages} {342} (\bibinfo {year} {2016})}\BibitemShut
  {NoStop}%
\bibitem [{\citenamefont {Wang}(2018)}]{Wang:2018csg}%
  \BibitemOpen
  \bibfield  {author} {\bibinfo {author} {\bibfnamefont {B.}~\bibnamefont
  {Wang}},\ }in\ \href@noop {} {\emph {\bibinfo {booktitle} {{36th
  International Symposium on Lattice Field Theory (Lattice 2018) East Lansing,
  MI, United States, July 22-28, 2018}}}}\ (\bibinfo {year} {2018})\ \Eprint
  {http://arxiv.org/abs/1812.05302} {arXiv:1812.05302 [hep-lat]} \BibitemShut
  {NoStop}%
\bibitem [{\citenamefont {Feng}\ \emph {et~al.}(2015)\citenamefont {Feng},
  \citenamefont {Christ}, \citenamefont {Portelli},\ and\ \citenamefont
  {Sachrajda}}]{Feng:2015kfa}%
  \BibitemOpen
  \bibfield  {author} {\bibinfo {author} {\bibfnamefont {X.}~\bibnamefont
  {Feng}}, \bibinfo {author} {\bibfnamefont {N.~H.}\ \bibnamefont {Christ}},
  \bibinfo {author} {\bibfnamefont {A.}~\bibnamefont {Portelli}}, \ and\
  \bibinfo {author} {\bibfnamefont {C.}~\bibnamefont {Sachrajda}},\ }\bibfield
  {booktitle} {\emph {\bibinfo {booktitle} {{Proceedings, 32nd International
  Symposium on Lattice Field Theory (Lattice 2014): Brookhaven, NY, USA, June
  23-28, 2014}}},\ }\href@noop {} {\bibfield  {journal} {\bibinfo  {journal}
  {PoS}\ }\textbf {\bibinfo {volume} {LATTICE2014}},\ \bibinfo {pages} {367}
  (\bibinfo {year} {2015})}\BibitemShut {NoStop}%
\bibitem [{\citenamefont {Christ}\ \emph
  {et~al.}(2016{\natexlab{a}})\citenamefont {Christ}, \citenamefont {Feng},
  \citenamefont {Portelli},\ and\ \citenamefont {Sachrajda}}]{Christ:2016eae}%
  \BibitemOpen
  \bibfield  {author} {\bibinfo {author} {\bibfnamefont {N.~H.}\ \bibnamefont
  {Christ}}, \bibinfo {author} {\bibfnamefont {X.}~\bibnamefont {Feng}},
  \bibinfo {author} {\bibfnamefont {A.}~\bibnamefont {Portelli}}, \ and\
  \bibinfo {author} {\bibfnamefont {C.~T.}\ \bibnamefont {Sachrajda}} (\bibinfo
  {collaboration} {RBC, UKQCD}),\ }\href {\doibase 10.1103/PhysRevD.93.114517}
  {\bibfield  {journal} {\bibinfo  {journal} {Phys. Rev.}\ }\textbf {\bibinfo
  {volume} {D93}},\ \bibinfo {pages} {114517} (\bibinfo {year}
  {2016}{\natexlab{a}})},\ \Eprint {http://arxiv.org/abs/1605.04442}
  {arXiv:1605.04442 [hep-lat]} \BibitemShut {NoStop}%
\bibitem [{\citenamefont {Christ}\ \emph
  {et~al.}(2016{\natexlab{b}})\citenamefont {Christ}, \citenamefont {Feng},
  \citenamefont {Lawson}, \citenamefont {Portelli},\ and\ \citenamefont
  {Sachrajda}}]{Christ:2016lro}%
  \BibitemOpen
  \bibfield  {author} {\bibinfo {author} {\bibfnamefont {N.~H.}\ \bibnamefont
  {Christ}}, \bibinfo {author} {\bibfnamefont {X.}~\bibnamefont {Feng}},
  \bibinfo {author} {\bibfnamefont {A.}~\bibnamefont {Lawson}}, \bibinfo
  {author} {\bibfnamefont {A.}~\bibnamefont {Portelli}}, \ and\ \bibinfo
  {author} {\bibfnamefont {C.}~\bibnamefont {Sachrajda}},\ }\bibfield
  {booktitle} {\emph {\bibinfo {booktitle} {{Proceedings, 34th International
  Symposium on Lattice Field Theory (Lattice 2016): Southampton, UK, July
  24-30, 2016}}},\ }\href@noop {} {\bibfield  {journal} {\bibinfo  {journal}
  {PoS}\ }\textbf {\bibinfo {volume} {LATTICE2016}},\ \bibinfo {pages} {306}
  (\bibinfo {year} {2016}{\natexlab{b}})}\BibitemShut {NoStop}%
\bibitem [{\citenamefont {Christ}\ \emph
  {et~al.}(2016{\natexlab{c}})\citenamefont {Christ}, \citenamefont {Feng},
  \citenamefont {Jüttner}, \citenamefont {Lawson}, \citenamefont {Portelli},\
  and\ \citenamefont {Sachrajda}}]{Christ:2016psm}%
  \BibitemOpen
  \bibfield  {author} {\bibinfo {author} {\bibfnamefont {N.~H.}\ \bibnamefont
  {Christ}}, \bibinfo {author} {\bibfnamefont {X.}~\bibnamefont {Feng}},
  \bibinfo {author} {\bibfnamefont {A.}~\bibnamefont {Jüttner}}, \bibinfo
  {author} {\bibfnamefont {A.}~\bibnamefont {Lawson}}, \bibinfo {author}
  {\bibfnamefont {A.}~\bibnamefont {Portelli}}, \ and\ \bibinfo {author}
  {\bibfnamefont {C.~T.}\ \bibnamefont {Sachrajda}},\ }\bibfield  {booktitle}
  {\emph {\bibinfo {booktitle} {{Proceedings, 8th International Workshop on
  Chiral Dynamics (CD15): Pisa, Italy, June 29-July 3, 2015}}},\ }\href@noop {}
  {\bibfield  {journal} {\bibinfo  {journal} {PoS}\ }\textbf {\bibinfo {volume}
  {CD15}},\ \bibinfo {pages} {033} (\bibinfo {year}
  {2016}{\natexlab{c}})}\BibitemShut {NoStop}%
\bibitem [{\citenamefont {Bai}\ \emph {et~al.}(2017)\citenamefont {Bai},
  \citenamefont {Christ}, \citenamefont {Feng}, \citenamefont {Lawson},
  \citenamefont {Portelli},\ and\ \citenamefont {Sachrajda}}]{Bai:2017fkh}%
  \BibitemOpen
  \bibfield  {author} {\bibinfo {author} {\bibfnamefont {Z.}~\bibnamefont
  {Bai}}, \bibinfo {author} {\bibfnamefont {N.~H.}\ \bibnamefont {Christ}},
  \bibinfo {author} {\bibfnamefont {X.}~\bibnamefont {Feng}}, \bibinfo {author}
  {\bibfnamefont {A.}~\bibnamefont {Lawson}}, \bibinfo {author} {\bibfnamefont
  {A.}~\bibnamefont {Portelli}}, \ and\ \bibinfo {author} {\bibfnamefont
  {C.~T.}\ \bibnamefont {Sachrajda}},\ }\href {\doibase
  10.1103/PhysRevLett.118.252001} {\bibfield  {journal} {\bibinfo  {journal}
  {Phys. Rev. Lett.}\ }\textbf {\bibinfo {volume} {118}},\ \bibinfo {pages}
  {252001} (\bibinfo {year} {2017})},\ \Eprint
  {http://arxiv.org/abs/1701.02858} {arXiv:1701.02858 [hep-lat]} \BibitemShut
  {NoStop}%
\bibitem [{\citenamefont {Bai}\ \emph {et~al.}(2018)\citenamefont {Bai},
  \citenamefont {Christ}, \citenamefont {Feng}, \citenamefont {Lawson},
  \citenamefont {Portelli},\ and\ \citenamefont {Sachrajda}}]{Bai:2018hqu}%
  \BibitemOpen
  \bibfield  {author} {\bibinfo {author} {\bibfnamefont {Z.}~\bibnamefont
  {Bai}}, \bibinfo {author} {\bibfnamefont {N.~H.}\ \bibnamefont {Christ}},
  \bibinfo {author} {\bibfnamefont {X.}~\bibnamefont {Feng}}, \bibinfo {author}
  {\bibfnamefont {A.}~\bibnamefont {Lawson}}, \bibinfo {author} {\bibfnamefont
  {A.}~\bibnamefont {Portelli}}, \ and\ \bibinfo {author} {\bibfnamefont
  {C.~T.}\ \bibnamefont {Sachrajda}},\ }\href {\doibase
  10.1103/PhysRevD.98.074509} {\bibfield  {journal} {\bibinfo  {journal} {Phys.
  Rev. D}\ }\textbf {\bibinfo {volume} {98}},\ \bibinfo {pages} {074509}
  (\bibinfo {year} {2018})},\ \Eprint {http://arxiv.org/abs/1806.11520}
  {arXiv:1806.11520 [hep-lat]} \BibitemShut {NoStop}%
\bibitem [{\citenamefont {Christ}\ \emph {et~al.}(2019)\citenamefont {Christ},
  \citenamefont {Feng}, \citenamefont {Portelli},\ and\ \citenamefont
  {Sachrajda}}]{Christ:2019dxu}%
  \BibitemOpen
  \bibfield  {author} {\bibinfo {author} {\bibfnamefont {N.~H.}\ \bibnamefont
  {Christ}}, \bibinfo {author} {\bibfnamefont {X.}~\bibnamefont {Feng}},
  \bibinfo {author} {\bibfnamefont {A.}~\bibnamefont {Portelli}}, \ and\
  \bibinfo {author} {\bibfnamefont {C.~T.}\ \bibnamefont {Sachrajda}} (\bibinfo
  {collaboration} {RBC, UKQCD}),\ }\href {\doibase 10.1103/PhysRevD.100.114506}
  {\bibfield  {journal} {\bibinfo  {journal} {Phys. Rev. D}\ }\textbf {\bibinfo
  {volume} {100}},\ \bibinfo {pages} {114506} (\bibinfo {year} {2019})},\
  \Eprint {http://arxiv.org/abs/1910.10644} {arXiv:1910.10644 [hep-lat]}
  \BibitemShut {NoStop}%
\bibitem [{\citenamefont {Christ}\ \emph
  {et~al.}(2015{\natexlab{a}})\citenamefont {Christ}, \citenamefont {Feng},
  \citenamefont {Portelli},\ and\ \citenamefont {Sachrajda}}]{Christ:2015aha}%
  \BibitemOpen
  \bibfield  {author} {\bibinfo {author} {\bibfnamefont {N.~H.}\ \bibnamefont
  {Christ}}, \bibinfo {author} {\bibfnamefont {X.}~\bibnamefont {Feng}},
  \bibinfo {author} {\bibfnamefont {A.}~\bibnamefont {Portelli}}, \ and\
  \bibinfo {author} {\bibfnamefont {C.~T.}\ \bibnamefont {Sachrajda}} (\bibinfo
  {collaboration} {RBC, UKQCD}),\ }\href {\doibase 10.1103/PhysRevD.92.094512}
  {\bibfield  {journal} {\bibinfo  {journal} {Phys. Rev.}\ }\textbf {\bibinfo
  {volume} {D92}},\ \bibinfo {pages} {094512} (\bibinfo {year}
  {2015}{\natexlab{a}})},\ \Eprint {http://arxiv.org/abs/1507.03094}
  {arXiv:1507.03094 [hep-lat]} \BibitemShut {NoStop}%
\bibitem [{\citenamefont {Christ}\ \emph
  {et~al.}(2016{\natexlab{d}})\citenamefont {Christ}, \citenamefont {Feng},
  \citenamefont {Juttner}, \citenamefont {Lawson}, \citenamefont {Portelli},\
  and\ \citenamefont {Sachrajda}}]{Christ:2016awg}%
  \BibitemOpen
  \bibfield  {author} {\bibinfo {author} {\bibfnamefont {N.}~\bibnamefont
  {Christ}}, \bibinfo {author} {\bibfnamefont {X.}~\bibnamefont {Feng}},
  \bibinfo {author} {\bibfnamefont {A.}~\bibnamefont {Juttner}}, \bibinfo
  {author} {\bibfnamefont {A.}~\bibnamefont {Lawson}}, \bibinfo {author}
  {\bibfnamefont {A.}~\bibnamefont {Portelli}}, \ and\ \bibinfo {author}
  {\bibfnamefont {C.}~\bibnamefont {Sachrajda}},\ }in\ \href
  {http://inspirehep.net/record/1419253/files/arXiv:1602.01374.pdf} {\emph
  {\bibinfo {booktitle} {{Proceedings, 33rd International Symposium on Lattice
  Field Theory (Lattice 2015)}}}}\ (\bibinfo {year} {2016})\ \Eprint
  {http://arxiv.org/abs/1602.01374} {arXiv:1602.01374 [hep-lat]} \BibitemShut
  {NoStop}%
\bibitem [{\citenamefont {Christ}\ \emph
  {et~al.}(2016{\natexlab{e}})\citenamefont {Christ}, \citenamefont {Feng},
  \citenamefont {Juttner}, \citenamefont {Lawson}, \citenamefont {Portelli},\
  and\ \citenamefont {Sachrajda}}]{Christ:2016mmq}%
  \BibitemOpen
  \bibfield  {author} {\bibinfo {author} {\bibfnamefont {N.~H.}\ \bibnamefont
  {Christ}}, \bibinfo {author} {\bibfnamefont {X.}~\bibnamefont {Feng}},
  \bibinfo {author} {\bibfnamefont {A.}~\bibnamefont {Juttner}}, \bibinfo
  {author} {\bibfnamefont {A.}~\bibnamefont {Lawson}}, \bibinfo {author}
  {\bibfnamefont {A.}~\bibnamefont {Portelli}}, \ and\ \bibinfo {author}
  {\bibfnamefont {C.~T.}\ \bibnamefont {Sachrajda}},\ }\href {\doibase
  10.1103/PhysRevD.94.114516} {\bibfield  {journal} {\bibinfo  {journal} {Phys.
  Rev.}\ }\textbf {\bibinfo {volume} {D94}},\ \bibinfo {pages} {114516}
  (\bibinfo {year} {2016}{\natexlab{e}})},\ \Eprint
  {http://arxiv.org/abs/1608.07585} {arXiv:1608.07585 [hep-lat]} \BibitemShut
  {NoStop}%
\bibitem [{\citenamefont {Lawson}\ \emph {et~al.}(2017)\citenamefont {Lawson},
  \citenamefont {Christ}, \citenamefont {Feng}, \citenamefont {Jüttner},
  \citenamefont {Portelli},\ and\ \citenamefont {Sachrajda}}]{Lawson:2017kxc}%
  \BibitemOpen
  \bibfield  {author} {\bibinfo {author} {\bibfnamefont {A.}~\bibnamefont
  {Lawson}}, \bibinfo {author} {\bibfnamefont {N.~H.}\ \bibnamefont {Christ}},
  \bibinfo {author} {\bibfnamefont {X.}~\bibnamefont {Feng}}, \bibinfo {author}
  {\bibfnamefont {A.}~\bibnamefont {Jüttner}}, \bibinfo {author}
  {\bibfnamefont {A.}~\bibnamefont {Portelli}}, \ and\ \bibinfo {author}
  {\bibfnamefont {C.}~\bibnamefont {Sachrajda}},\ }\bibfield  {booktitle}
  {\emph {\bibinfo {booktitle} {{Proceedings, 34th International Symposium on
  Lattice Field Theory (Lattice 2016): Southampton, UK, July 24-30, 2016}}},\
  }\href {\doibase 10.22323/1.256.0303} {\bibfield  {journal} {\bibinfo
  {journal} {PoS}\ }\textbf {\bibinfo {volume} {LATTICE2016}},\ \bibinfo
  {pages} {303} (\bibinfo {year} {2017})}\BibitemShut {NoStop}%
\bibitem [{\citenamefont {Tiburzi}\ \emph {et~al.}(2017)\citenamefont
  {Tiburzi}, \citenamefont {Wagman}, \citenamefont {Winter}, \citenamefont
  {Chang}, \citenamefont {Davoudi}, \citenamefont {Detmold}, \citenamefont
  {Orginos}, \citenamefont {Savage},\ and\ \citenamefont
  {Shanahan}}]{Tiburzi:2017iux}%
  \BibitemOpen
  \bibfield  {author} {\bibinfo {author} {\bibfnamefont {B.~C.}\ \bibnamefont
  {Tiburzi}}, \bibinfo {author} {\bibfnamefont {M.~L.}\ \bibnamefont {Wagman}},
  \bibinfo {author} {\bibfnamefont {F.}~\bibnamefont {Winter}}, \bibinfo
  {author} {\bibfnamefont {E.}~\bibnamefont {Chang}}, \bibinfo {author}
  {\bibfnamefont {Z.}~\bibnamefont {Davoudi}}, \bibinfo {author} {\bibfnamefont
  {W.}~\bibnamefont {Detmold}}, \bibinfo {author} {\bibfnamefont
  {K.}~\bibnamefont {Orginos}}, \bibinfo {author} {\bibfnamefont {M.~J.}\
  \bibnamefont {Savage}}, \ and\ \bibinfo {author} {\bibfnamefont {P.~E.}\
  \bibnamefont {Shanahan}},\ }\href {\doibase 10.1103/PhysRevD.96.054505}
  {\bibfield  {journal} {\bibinfo  {journal} {Phys. Rev.}\ }\textbf {\bibinfo
  {volume} {D96}},\ \bibinfo {pages} {054505} (\bibinfo {year} {2017})},\
  \Eprint {http://arxiv.org/abs/1702.02929} {arXiv:1702.02929 [hep-lat]}
  \BibitemShut {NoStop}%
\bibitem [{\citenamefont {Shanahan}\ \emph {et~al.}(2017)\citenamefont
  {Shanahan}, \citenamefont {Tiburzi}, \citenamefont {Wagman}, \citenamefont
  {Winter}, \citenamefont {Chang}, \citenamefont {Davoudi}, \citenamefont
  {Detmold}, \citenamefont {Orginos},\ and\ \citenamefont
  {Savage}}]{Shanahan:2017bgi}%
  \BibitemOpen
  \bibfield  {author} {\bibinfo {author} {\bibfnamefont {P.~E.}\ \bibnamefont
  {Shanahan}}, \bibinfo {author} {\bibfnamefont {B.~C.}\ \bibnamefont
  {Tiburzi}}, \bibinfo {author} {\bibfnamefont {M.~L.}\ \bibnamefont {Wagman}},
  \bibinfo {author} {\bibfnamefont {F.}~\bibnamefont {Winter}}, \bibinfo
  {author} {\bibfnamefont {E.}~\bibnamefont {Chang}}, \bibinfo {author}
  {\bibfnamefont {Z.}~\bibnamefont {Davoudi}}, \bibinfo {author} {\bibfnamefont
  {W.}~\bibnamefont {Detmold}}, \bibinfo {author} {\bibfnamefont
  {K.}~\bibnamefont {Orginos}}, \ and\ \bibinfo {author} {\bibfnamefont
  {M.~J.}\ \bibnamefont {Savage}},\ }\href {\doibase
  10.1103/PhysRevLett.119.062003} {\bibfield  {journal} {\bibinfo  {journal}
  {Phys. Rev. Lett.}\ }\textbf {\bibinfo {volume} {119}},\ \bibinfo {pages}
  {062003} (\bibinfo {year} {2017})},\ \Eprint
  {http://arxiv.org/abs/1701.03456} {arXiv:1701.03456 [hep-lat]} \BibitemShut
  {NoStop}%
\bibitem [{\citenamefont {Nicholson}\ \emph
  {et~al.}(2018{\natexlab{a}})\citenamefont {Nicholson} \emph
  {et~al.}}]{Nicholson:2018mwc}%
  \BibitemOpen
  \bibfield  {author} {\bibinfo {author} {\bibfnamefont {A.}~\bibnamefont
  {Nicholson}} \emph {et~al.},\ }\href {\doibase
  10.1103/PhysRevLett.121.172501} {\bibfield  {journal} {\bibinfo  {journal}
  {Phys. Rev. Lett.}\ }\textbf {\bibinfo {volume} {121}},\ \bibinfo {pages}
  {172501} (\bibinfo {year} {2018}{\natexlab{a}})},\ \Eprint
  {http://arxiv.org/abs/1805.02634} {arXiv:1805.02634 [nucl-th]} \BibitemShut
  {NoStop}%
\bibitem [{\citenamefont {Feng}\ \emph {et~al.}(2019)\citenamefont {Feng},
  \citenamefont {Jin}, \citenamefont {Tuo},\ and\ \citenamefont
  {Xia}}]{Feng:2018pdq}%
  \BibitemOpen
  \bibfield  {author} {\bibinfo {author} {\bibfnamefont {X.}~\bibnamefont
  {Feng}}, \bibinfo {author} {\bibfnamefont {L.-C.}\ \bibnamefont {Jin}},
  \bibinfo {author} {\bibfnamefont {X.-Y.}\ \bibnamefont {Tuo}}, \ and\
  \bibinfo {author} {\bibfnamefont {S.-C.}\ \bibnamefont {Xia}},\ }\href
  {\doibase 10.1103/PhysRevLett.122.022001} {\bibfield  {journal} {\bibinfo
  {journal} {Phys. Rev. Lett.}\ }\textbf {\bibinfo {volume} {122}},\ \bibinfo
  {pages} {022001} (\bibinfo {year} {2019})},\ \Eprint
  {http://arxiv.org/abs/1809.10511} {arXiv:1809.10511 [hep-lat]} \BibitemShut
  {NoStop}%
\bibitem [{\citenamefont {Nicholson}\ \emph
  {et~al.}(2018{\natexlab{b}})\citenamefont {Nicholson} \emph
  {et~al.}}]{Nicholson:2018laj}%
  \BibitemOpen
  \bibfield  {author} {\bibinfo {author} {\bibfnamefont {A.}~\bibnamefont
  {Nicholson}} \emph {et~al.}\ }(\bibinfo {year} {2018})\ \Eprint
  {http://arxiv.org/abs/1812.11127} {arXiv:1812.11127 [hep-lat]} \BibitemShut
  {NoStop}%
\bibitem [{\citenamefont {Tuo}\ \emph {et~al.}(2019)\citenamefont {Tuo},
  \citenamefont {Feng},\ and\ \citenamefont {Jin}}]{Tuo:2019bue}%
  \BibitemOpen
  \bibfield  {author} {\bibinfo {author} {\bibfnamefont {X.-Y.}\ \bibnamefont
  {Tuo}}, \bibinfo {author} {\bibfnamefont {X.}~\bibnamefont {Feng}}, \ and\
  \bibinfo {author} {\bibfnamefont {L.-C.}\ \bibnamefont {Jin}},\ }\href
  {\doibase 10.1103/PhysRevD.100.094511} {\bibfield  {journal} {\bibinfo
  {journal} {Phys. Rev. D}\ }\textbf {\bibinfo {volume} {100}},\ \bibinfo
  {pages} {094511} (\bibinfo {year} {2019})},\ \Eprint
  {http://arxiv.org/abs/1909.13525} {arXiv:1909.13525 [hep-lat]} \BibitemShut
  {NoStop}%
\bibitem [{\citenamefont {Cirigliano}\ \emph {et~al.}(2020)\citenamefont
  {Cirigliano}, \citenamefont {Detmold}, \citenamefont {Nicholson},\ and\
  \citenamefont {Shanahan}}]{Cirigliano:2020yhp}%
  \BibitemOpen
  \bibfield  {author} {\bibinfo {author} {\bibfnamefont {V.}~\bibnamefont
  {Cirigliano}}, \bibinfo {author} {\bibfnamefont {W.}~\bibnamefont {Detmold}},
  \bibinfo {author} {\bibfnamefont {A.}~\bibnamefont {Nicholson}}, \ and\
  \bibinfo {author} {\bibfnamefont {P.}~\bibnamefont {Shanahan}},\ }\href
  {\doibase 10.1016/j.ppnp.2020.103771} {\  (\bibinfo {year} {2020}),\
  10.1016/j.ppnp.2020.103771},\ \Eprint {http://arxiv.org/abs/2003.08493}
  {arXiv:2003.08493 [nucl-th]} \BibitemShut {NoStop}%
\bibitem [{\citenamefont {Detmold}\ and\ \citenamefont
  {Murphy}(2020)}]{Detmold:2020jqv}%
  \BibitemOpen
  \bibfield  {author} {\bibinfo {author} {\bibfnamefont {W.}~\bibnamefont
  {Detmold}}\ and\ \bibinfo {author} {\bibfnamefont {D.}~\bibnamefont {Murphy}}
  (\bibinfo {collaboration} {NPLQCD}),\ }\href@noop {} {\  (\bibinfo {year}
  {2020})},\ \Eprint {http://arxiv.org/abs/2004.07404} {arXiv:2004.07404
  [hep-lat]} \BibitemShut {NoStop}%
\bibitem [{\citenamefont {Feng}\ \emph {et~al.}(2020)\citenamefont {Feng},
  \citenamefont {Gorchtein}, \citenamefont {Jin}, \citenamefont {Ma},\ and\
  \citenamefont {Seng}}]{Feng:2020zdc}%
  \BibitemOpen
  \bibfield  {author} {\bibinfo {author} {\bibfnamefont {X.}~\bibnamefont
  {Feng}}, \bibinfo {author} {\bibfnamefont {M.}~\bibnamefont {Gorchtein}},
  \bibinfo {author} {\bibfnamefont {L.-C.}\ \bibnamefont {Jin}}, \bibinfo
  {author} {\bibfnamefont {P.-X.}\ \bibnamefont {Ma}}, \ and\ \bibinfo {author}
  {\bibfnamefont {C.-Y.}\ \bibnamefont {Seng}},\ }\href {\doibase
  10.1103/PhysRevLett.124.192002} {\bibfield  {journal} {\bibinfo  {journal}
  {Phys. Rev. Lett.}\ }\textbf {\bibinfo {volume} {124}},\ \bibinfo {pages}
  {192002} (\bibinfo {year} {2020})},\ \Eprint
  {http://arxiv.org/abs/2003.09798} {arXiv:2003.09798 [hep-lat]} \BibitemShut
  {NoStop}%
\bibitem [{\citenamefont {Hashimoto}(2017)}]{Hashimoto:2017wqo}%
  \BibitemOpen
  \bibfield  {author} {\bibinfo {author} {\bibfnamefont {S.}~\bibnamefont
  {Hashimoto}},\ }\href {\doibase 10.1093/ptep/ptx052} {\bibfield  {journal}
  {\bibinfo  {journal} {PTEP}\ }\textbf {\bibinfo {volume} {2017}},\ \bibinfo
  {pages} {053B03} (\bibinfo {year} {2017})},\ \Eprint
  {http://arxiv.org/abs/1703.01881} {arXiv:1703.01881 [hep-lat]} \BibitemShut
  {NoStop}%
\bibitem [{\citenamefont {Hashimoto}(2018)}]{Hashimoto:2019pgh}%
  \BibitemOpen
  \bibfield  {author} {\bibinfo {author} {\bibfnamefont {S.}~\bibnamefont
  {Hashimoto}},\ }\href {\doibase 10.22323/1.334.0008} {\bibfield  {journal}
  {\bibinfo  {journal} {PoS}\ }\textbf {\bibinfo {volume} {LATTICE2018}},\
  \bibinfo {pages} {008} (\bibinfo {year} {2018})},\ \Eprint
  {http://arxiv.org/abs/1902.09119} {arXiv:1902.09119 [hep-lat]} \BibitemShut
  {NoStop}%
\bibitem [{\citenamefont {Gambino}\ and\ \citenamefont
  {Hashimoto}(2020)}]{Gambino:2020crt}%
  \BibitemOpen
  \bibfield  {author} {\bibinfo {author} {\bibfnamefont {P.}~\bibnamefont
  {Gambino}}\ and\ \bibinfo {author} {\bibfnamefont {S.}~\bibnamefont
  {Hashimoto}},\ }\href {\doibase 10.1103/PhysRevLett.125.032001} {\bibfield
  {journal} {\bibinfo  {journal} {Phys. Rev. Lett.}\ }\textbf {\bibinfo
  {volume} {125}},\ \bibinfo {pages} {032001} (\bibinfo {year} {2020})},\
  \Eprint {http://arxiv.org/abs/2005.13730} {arXiv:2005.13730 [hep-lat]}
  \BibitemShut {NoStop}%
\bibitem [{\citenamefont {Chambers}\ \emph {et~al.}(2017)\citenamefont
  {Chambers}, \citenamefont {Horsley}, \citenamefont {Nakamura}, \citenamefont
  {Perlt}, \citenamefont {Rakow}, \citenamefont {Schierholz}, \citenamefont
  {Schiller}, \citenamefont {Somfleth}, \citenamefont {Young},\ and\
  \citenamefont {Zanotti}}]{Chambers:2017dov}%
  \BibitemOpen
  \bibfield  {author} {\bibinfo {author} {\bibfnamefont {A.~J.}\ \bibnamefont
  {Chambers}}, \bibinfo {author} {\bibfnamefont {R.}~\bibnamefont {Horsley}},
  \bibinfo {author} {\bibfnamefont {Y.}~\bibnamefont {Nakamura}}, \bibinfo
  {author} {\bibfnamefont {H.}~\bibnamefont {Perlt}}, \bibinfo {author}
  {\bibfnamefont {P.~E.~L.}\ \bibnamefont {Rakow}}, \bibinfo {author}
  {\bibfnamefont {G.}~\bibnamefont {Schierholz}}, \bibinfo {author}
  {\bibfnamefont {A.}~\bibnamefont {Schiller}}, \bibinfo {author}
  {\bibfnamefont {K.}~\bibnamefont {Somfleth}}, \bibinfo {author}
  {\bibfnamefont {R.~D.}\ \bibnamefont {Young}}, \ and\ \bibinfo {author}
  {\bibfnamefont {J.~M.}\ \bibnamefont {Zanotti}},\ }\href {\doibase
  10.1103/PhysRevLett.118.242001} {\bibfield  {journal} {\bibinfo  {journal}
  {Phys. Rev. Lett.}\ }\textbf {\bibinfo {volume} {118}},\ \bibinfo {pages}
  {242001} (\bibinfo {year} {2017})},\ \Eprint
  {http://arxiv.org/abs/1703.01153} {arXiv:1703.01153 [hep-lat]} \BibitemShut
  {NoStop}%
\bibitem [{\citenamefont {Liu}(2017)}]{Liu:2017lpe}%
  \BibitemOpen
  \bibfield  {author} {\bibinfo {author} {\bibfnamefont {K.-F.}\ \bibnamefont
  {Liu}},\ }\href {\doibase 10.1103/PhysRevD.96.033001} {\bibfield  {journal}
  {\bibinfo  {journal} {Phys. Rev.}\ }\textbf {\bibinfo {volume} {D96}},\
  \bibinfo {pages} {033001} (\bibinfo {year} {2017})},\ \Eprint
  {http://arxiv.org/abs/1703.04690} {arXiv:1703.04690 [hep-ph]} \BibitemShut
  {NoStop}%
\bibitem [{\citenamefont {Hansen}\ \emph {et~al.}(2017)\citenamefont {Hansen},
  \citenamefont {Meyer},\ and\ \citenamefont {Robaina}}]{Hansen:2017mnd}%
  \BibitemOpen
  \bibfield  {author} {\bibinfo {author} {\bibfnamefont {M.~T.}\ \bibnamefont
  {Hansen}}, \bibinfo {author} {\bibfnamefont {H.~B.}\ \bibnamefont {Meyer}}, \
  and\ \bibinfo {author} {\bibfnamefont {D.}~\bibnamefont {Robaina}},\ }\href
  {\doibase 10.1103/PhysRevD.96.094513} {\bibfield  {journal} {\bibinfo
  {journal} {Phys. Rev.}\ }\textbf {\bibinfo {volume} {D96}},\ \bibinfo {pages}
  {094513} (\bibinfo {year} {2017})},\ \Eprint
  {http://arxiv.org/abs/1704.08993} {arXiv:1704.08993 [hep-lat]} \BibitemShut
  {NoStop}%
\bibitem [{\citenamefont {Can}\ \emph {et~al.}(2020)\citenamefont {Can} \emph
  {et~al.}}]{Can:2020sxc}%
  \BibitemOpen
  \bibfield  {author} {\bibinfo {author} {\bibfnamefont {K.}~\bibnamefont
  {Can}} \emph {et~al.},\ }\href@noop {} {\  (\bibinfo {year} {2020})},\
  \Eprint {http://arxiv.org/abs/2007.01523} {arXiv:2007.01523 [hep-lat]}
  \BibitemShut {NoStop}%
\bibitem [{\citenamefont {Duncan}\ \emph {et~al.}(1996)\citenamefont {Duncan},
  \citenamefont {Eichten},\ and\ \citenamefont {Thacker}}]{Duncan:1996xy}%
  \BibitemOpen
  \bibfield  {author} {\bibinfo {author} {\bibfnamefont {A.}~\bibnamefont
  {Duncan}}, \bibinfo {author} {\bibfnamefont {E.}~\bibnamefont {Eichten}}, \
  and\ \bibinfo {author} {\bibfnamefont {H.}~\bibnamefont {Thacker}},\ }\href
  {\doibase 10.1103/PhysRevLett.76.3894} {\bibfield  {journal} {\bibinfo
  {journal} {Phys. Rev. Lett.}\ }\textbf {\bibinfo {volume} {76}},\ \bibinfo
  {pages} {3894} (\bibinfo {year} {1996})},\ \Eprint
  {http://arxiv.org/abs/hep-lat/9602005} {arXiv:hep-lat/9602005 [hep-lat]}
  \BibitemShut {NoStop}%
\bibitem [{\citenamefont {Duncan}\ \emph {et~al.}(1997)\citenamefont {Duncan},
  \citenamefont {Eichten},\ and\ \citenamefont {Thacker}}]{Duncan:1996be}%
  \BibitemOpen
  \bibfield  {author} {\bibinfo {author} {\bibfnamefont {A.}~\bibnamefont
  {Duncan}}, \bibinfo {author} {\bibfnamefont {E.}~\bibnamefont {Eichten}}, \
  and\ \bibinfo {author} {\bibfnamefont {H.}~\bibnamefont {Thacker}},\ }\href
  {\doibase 10.1016/S0370-2693(97)00850-2} {\bibfield  {journal} {\bibinfo
  {journal} {Phys. Lett.}\ }\textbf {\bibinfo {volume} {B409}},\ \bibinfo
  {pages} {387} (\bibinfo {year} {1997})},\ \Eprint
  {http://arxiv.org/abs/hep-lat/9607032} {arXiv:hep-lat/9607032 [hep-lat]}
  \BibitemShut {NoStop}%
\bibitem [{\citenamefont {Blum}\ \emph {et~al.}(2007)\citenamefont {Blum},
  \citenamefont {Doi}, \citenamefont {Hayakawa}, \citenamefont {Izubuchi},\
  and\ \citenamefont {Yamada}}]{Blum:2007cy}%
  \BibitemOpen
  \bibfield  {author} {\bibinfo {author} {\bibfnamefont {T.}~\bibnamefont
  {Blum}}, \bibinfo {author} {\bibfnamefont {T.}~\bibnamefont {Doi}}, \bibinfo
  {author} {\bibfnamefont {M.}~\bibnamefont {Hayakawa}}, \bibinfo {author}
  {\bibfnamefont {T.}~\bibnamefont {Izubuchi}}, \ and\ \bibinfo {author}
  {\bibfnamefont {N.}~\bibnamefont {Yamada}},\ }\href {\doibase
  10.1103/PhysRevD.76.114508} {\bibfield  {journal} {\bibinfo  {journal} {Phys.
  Rev.}\ }\textbf {\bibinfo {volume} {D76}},\ \bibinfo {pages} {114508}
  (\bibinfo {year} {2007})},\ \Eprint {http://arxiv.org/abs/0708.0484}
  {arXiv:0708.0484 [hep-lat]} \BibitemShut {NoStop}%
\bibitem [{\citenamefont {Blum}\ \emph {et~al.}(2010)\citenamefont {Blum},
  \citenamefont {Zhou}, \citenamefont {Doi}, \citenamefont {Hayakawa},
  \citenamefont {Izubuchi}, \citenamefont {Uno},\ and\ \citenamefont
  {Yamada}}]{Blum:2010ym}%
  \BibitemOpen
  \bibfield  {author} {\bibinfo {author} {\bibfnamefont {T.}~\bibnamefont
  {Blum}}, \bibinfo {author} {\bibfnamefont {R.}~\bibnamefont {Zhou}}, \bibinfo
  {author} {\bibfnamefont {T.}~\bibnamefont {Doi}}, \bibinfo {author}
  {\bibfnamefont {M.}~\bibnamefont {Hayakawa}}, \bibinfo {author}
  {\bibfnamefont {T.}~\bibnamefont {Izubuchi}}, \bibinfo {author}
  {\bibfnamefont {S.}~\bibnamefont {Uno}}, \ and\ \bibinfo {author}
  {\bibfnamefont {N.}~\bibnamefont {Yamada}},\ }\href {\doibase
  10.1103/PhysRevD.82.094508} {\bibfield  {journal} {\bibinfo  {journal} {Phys.
  Rev.}\ }\textbf {\bibinfo {volume} {D82}},\ \bibinfo {pages} {094508}
  (\bibinfo {year} {2010})},\ \Eprint {http://arxiv.org/abs/1006.1311}
  {arXiv:1006.1311 [hep-lat]} \BibitemShut {NoStop}%
\bibitem [{\citenamefont {Ishikawa}\ \emph {et~al.}(2012)\citenamefont
  {Ishikawa}, \citenamefont {Blum}, \citenamefont {Hayakawa}, \citenamefont
  {Izubuchi}, \citenamefont {Jung},\ and\ \citenamefont
  {Zhou}}]{Ishikawa:2012ix}%
  \BibitemOpen
  \bibfield  {author} {\bibinfo {author} {\bibfnamefont {T.}~\bibnamefont
  {Ishikawa}}, \bibinfo {author} {\bibfnamefont {T.}~\bibnamefont {Blum}},
  \bibinfo {author} {\bibfnamefont {M.}~\bibnamefont {Hayakawa}}, \bibinfo
  {author} {\bibfnamefont {T.}~\bibnamefont {Izubuchi}}, \bibinfo {author}
  {\bibfnamefont {C.}~\bibnamefont {Jung}}, \ and\ \bibinfo {author}
  {\bibfnamefont {R.}~\bibnamefont {Zhou}},\ }\href {\doibase
  10.1103/PhysRevLett.109.072002} {\bibfield  {journal} {\bibinfo  {journal}
  {Phys. Rev. Lett.}\ }\textbf {\bibinfo {volume} {109}},\ \bibinfo {pages}
  {072002} (\bibinfo {year} {2012})},\ \Eprint {http://arxiv.org/abs/1202.6018}
  {arXiv:1202.6018 [hep-lat]} \BibitemShut {NoStop}%
\bibitem [{\citenamefont {de~Divitiis}\ \emph {et~al.}(2013)\citenamefont
  {de~Divitiis}, \citenamefont {Frezzotti}, \citenamefont {Lubicz},
  \citenamefont {Martinelli}, \citenamefont {Petronzio}, \citenamefont {Rossi},
  \citenamefont {Sanfilippo}, \citenamefont {Simula},\ and\ \citenamefont
  {Tantalo}}]{deDivitiis:2013xla}%
  \BibitemOpen
  \bibfield  {author} {\bibinfo {author} {\bibfnamefont {G.~M.}\ \bibnamefont
  {de~Divitiis}}, \bibinfo {author} {\bibfnamefont {R.}~\bibnamefont
  {Frezzotti}}, \bibinfo {author} {\bibfnamefont {V.}~\bibnamefont {Lubicz}},
  \bibinfo {author} {\bibfnamefont {G.}~\bibnamefont {Martinelli}}, \bibinfo
  {author} {\bibfnamefont {R.}~\bibnamefont {Petronzio}}, \bibinfo {author}
  {\bibfnamefont {G.~C.}\ \bibnamefont {Rossi}}, \bibinfo {author}
  {\bibfnamefont {F.}~\bibnamefont {Sanfilippo}}, \bibinfo {author}
  {\bibfnamefont {S.}~\bibnamefont {Simula}}, \ and\ \bibinfo {author}
  {\bibfnamefont {N.}~\bibnamefont {Tantalo}} (\bibinfo {collaboration}
  {RM123}),\ }\href {\doibase 10.1103/PhysRevD.87.114505} {\bibfield  {journal}
  {\bibinfo  {journal} {Phys. Rev.}\ }\textbf {\bibinfo {volume} {D87}},\
  \bibinfo {pages} {114505} (\bibinfo {year} {2013})},\ \Eprint
  {http://arxiv.org/abs/1303.4896} {arXiv:1303.4896 [hep-lat]} \BibitemShut
  {NoStop}%
\bibitem [{\citenamefont {Borsanyi}\ \emph {et~al.}(2015)\citenamefont
  {Borsanyi} \emph {et~al.}}]{Borsanyi:2014jba}%
  \BibitemOpen
  \bibfield  {author} {\bibinfo {author} {\bibfnamefont {S.}~\bibnamefont
  {Borsanyi}} \emph {et~al.},\ }\href {\doibase 10.1126/science.1257050}
  {\bibfield  {journal} {\bibinfo  {journal} {Science}\ }\textbf {\bibinfo
  {volume} {347}},\ \bibinfo {pages} {1452} (\bibinfo {year} {2015})},\ \Eprint
  {http://arxiv.org/abs/1406.4088} {arXiv:1406.4088 [hep-lat]} \BibitemShut
  {NoStop}%
\bibitem [{\citenamefont {Horsley}\ \emph {et~al.}(2016)\citenamefont {Horsley}
  \emph {et~al.}}]{Horsley:2015vla}%
  \BibitemOpen
  \bibfield  {author} {\bibinfo {author} {\bibfnamefont {R.}~\bibnamefont
  {Horsley}} \emph {et~al.},\ }\href {\doibase 10.1007/JHEP04(2016)093}
  {\bibfield  {journal} {\bibinfo  {journal} {JHEP}\ }\textbf {\bibinfo
  {volume} {04}},\ \bibinfo {pages} {093} (\bibinfo {year} {2016})},\ \Eprint
  {http://arxiv.org/abs/1509.00799} {arXiv:1509.00799 [hep-lat]} \BibitemShut
  {NoStop}%
\bibitem [{\citenamefont {Giusti}\ \emph
  {et~al.}(2017{\natexlab{a}})\citenamefont {Giusti}, \citenamefont {Lubicz},
  \citenamefont {Tarantino}, \citenamefont {Martinelli}, \citenamefont
  {Sanfilippo}, \citenamefont {Simula},\ and\ \citenamefont
  {Tantalo}}]{Giusti:2017dmp}%
  \BibitemOpen
  \bibfield  {author} {\bibinfo {author} {\bibfnamefont {D.}~\bibnamefont
  {Giusti}}, \bibinfo {author} {\bibfnamefont {V.}~\bibnamefont {Lubicz}},
  \bibinfo {author} {\bibfnamefont {C.}~\bibnamefont {Tarantino}}, \bibinfo
  {author} {\bibfnamefont {G.}~\bibnamefont {Martinelli}}, \bibinfo {author}
  {\bibfnamefont {S.}~\bibnamefont {Sanfilippo}}, \bibinfo {author}
  {\bibfnamefont {S.}~\bibnamefont {Simula}}, \ and\ \bibinfo {author}
  {\bibfnamefont {N.}~\bibnamefont {Tantalo}},\ }\href {\doibase
  10.1103/PhysRevD.95.114504} {\bibfield  {journal} {\bibinfo  {journal} {Phys.
  Rev.}\ }\textbf {\bibinfo {volume} {D95}},\ \bibinfo {pages} {114504}
  (\bibinfo {year} {2017}{\natexlab{a}})},\ \Eprint
  {http://arxiv.org/abs/1704.06561} {arXiv:1704.06561 [hep-lat]} \BibitemShut
  {NoStop}%
\bibitem [{\citenamefont {Davoudi}\ \emph {et~al.}(2019)\citenamefont
  {Davoudi}, \citenamefont {Harrison}, \citenamefont {Jüttner}, \citenamefont
  {Portelli},\ and\ \citenamefont {Savage}}]{Davoudi:2018qpl}%
  \BibitemOpen
  \bibfield  {author} {\bibinfo {author} {\bibfnamefont {Z.}~\bibnamefont
  {Davoudi}}, \bibinfo {author} {\bibfnamefont {J.}~\bibnamefont {Harrison}},
  \bibinfo {author} {\bibfnamefont {A.}~\bibnamefont {Jüttner}}, \bibinfo
  {author} {\bibfnamefont {A.}~\bibnamefont {Portelli}}, \ and\ \bibinfo
  {author} {\bibfnamefont {M.~J.}\ \bibnamefont {Savage}},\ }\href {\doibase
  10.1103/PhysRevD.99.034510} {\bibfield  {journal} {\bibinfo  {journal} {Phys.
  Rev. D}\ }\textbf {\bibinfo {volume} {99}},\ \bibinfo {pages} {034510}
  (\bibinfo {year} {2019})},\ \Eprint {http://arxiv.org/abs/1810.05923}
  {arXiv:1810.05923 [hep-lat]} \BibitemShut {NoStop}%
\bibitem [{\citenamefont {Giusti}\ \emph
  {et~al.}(2017{\natexlab{b}})\citenamefont {Giusti}, \citenamefont {Lubicz},
  \citenamefont {Martinelli}, \citenamefont {Sanfilippo},\ and\ \citenamefont
  {Simula}}]{Giusti:2017jof}%
  \BibitemOpen
  \bibfield  {author} {\bibinfo {author} {\bibfnamefont {D.}~\bibnamefont
  {Giusti}}, \bibinfo {author} {\bibfnamefont {V.}~\bibnamefont {Lubicz}},
  \bibinfo {author} {\bibfnamefont {G.}~\bibnamefont {Martinelli}}, \bibinfo
  {author} {\bibfnamefont {F.}~\bibnamefont {Sanfilippo}}, \ and\ \bibinfo
  {author} {\bibfnamefont {S.}~\bibnamefont {Simula}},\ }\href {\doibase
  10.1007/JHEP10(2017)157} {\bibfield  {journal} {\bibinfo  {journal} {JHEP}\
  }\textbf {\bibinfo {volume} {10}},\ \bibinfo {pages} {157} (\bibinfo {year}
  {2017}{\natexlab{b}})},\ \Eprint {http://arxiv.org/abs/1707.03019}
  {arXiv:1707.03019 [hep-lat]} \BibitemShut {NoStop}%
\bibitem [{\citenamefont {Chakraborty}\ \emph {et~al.}(2018)\citenamefont
  {Chakraborty} \emph {et~al.}}]{Chakraborty:2017tqp}%
  \BibitemOpen
  \bibfield  {author} {\bibinfo {author} {\bibfnamefont {B.}~\bibnamefont
  {Chakraborty}} \emph {et~al.} (\bibinfo {collaboration} {Fermilab Lattice,
  LATTICE-HPQCD, MILC}),\ }\href {\doibase 10.1103/PhysRevLett.120.152001}
  {\bibfield  {journal} {\bibinfo  {journal} {Phys. Rev. Lett.}\ }\textbf
  {\bibinfo {volume} {120}},\ \bibinfo {pages} {152001} (\bibinfo {year}
  {2018})},\ \Eprint {http://arxiv.org/abs/1710.11212} {arXiv:1710.11212
  [hep-lat]} \BibitemShut {NoStop}%
\bibitem [{\citenamefont {Boyle}\ \emph {et~al.}(2017)\citenamefont {Boyle},
  \citenamefont {Gülpers}, \citenamefont {Harrison}, \citenamefont {Jüttner},
  \citenamefont {Lehner}, \citenamefont {Portelli},\ and\ \citenamefont
  {Sachrajda}}]{Boyle:2017gzv}%
  \BibitemOpen
  \bibfield  {author} {\bibinfo {author} {\bibfnamefont {P.}~\bibnamefont
  {Boyle}}, \bibinfo {author} {\bibfnamefont {V.}~\bibnamefont {Gülpers}},
  \bibinfo {author} {\bibfnamefont {J.}~\bibnamefont {Harrison}}, \bibinfo
  {author} {\bibfnamefont {A.}~\bibnamefont {Jüttner}}, \bibinfo {author}
  {\bibfnamefont {C.}~\bibnamefont {Lehner}}, \bibinfo {author} {\bibfnamefont
  {A.}~\bibnamefont {Portelli}}, \ and\ \bibinfo {author} {\bibfnamefont
  {C.~T.}\ \bibnamefont {Sachrajda}},\ }\href {\doibase
  10.1007/JHEP09(2017)153} {\bibfield  {journal} {\bibinfo  {journal} {JHEP}\
  }\textbf {\bibinfo {volume} {09}},\ \bibinfo {pages} {153} (\bibinfo {year}
  {2017})},\ \Eprint {http://arxiv.org/abs/1706.05293} {arXiv:1706.05293
  [hep-lat]} \BibitemShut {NoStop}%
\bibitem [{\citenamefont {Blum}\ \emph {et~al.}(2018)\citenamefont {Blum},
  \citenamefont {Boyle}, \citenamefont {Gülpers}, \citenamefont {Izubuchi},
  \citenamefont {Jin}, \citenamefont {Jung}, \citenamefont {Jüttner},
  \citenamefont {Lehner}, \citenamefont {Portelli},\ and\ \citenamefont
  {Tsang}}]{Blum:2018mom}%
  \BibitemOpen
  \bibfield  {author} {\bibinfo {author} {\bibfnamefont {T.}~\bibnamefont
  {Blum}}, \bibinfo {author} {\bibfnamefont {P.~A.}\ \bibnamefont {Boyle}},
  \bibinfo {author} {\bibfnamefont {V.}~\bibnamefont {Gülpers}}, \bibinfo
  {author} {\bibfnamefont {T.}~\bibnamefont {Izubuchi}}, \bibinfo {author}
  {\bibfnamefont {L.}~\bibnamefont {Jin}}, \bibinfo {author} {\bibfnamefont
  {C.}~\bibnamefont {Jung}}, \bibinfo {author} {\bibfnamefont {A.}~\bibnamefont
  {Jüttner}}, \bibinfo {author} {\bibfnamefont {C.}~\bibnamefont {Lehner}},
  \bibinfo {author} {\bibfnamefont {A.}~\bibnamefont {Portelli}}, \ and\
  \bibinfo {author} {\bibfnamefont {J.~T.}\ \bibnamefont {Tsang}} (\bibinfo
  {collaboration} {RBC, UKQCD}),\ }\href {\doibase
  10.1103/PhysRevLett.121.022003} {\bibfield  {journal} {\bibinfo  {journal}
  {Phys. Rev. Lett.}\ }\textbf {\bibinfo {volume} {121}},\ \bibinfo {pages}
  {022003} (\bibinfo {year} {2018})},\ \Eprint
  {http://arxiv.org/abs/1801.07224} {arXiv:1801.07224 [hep-lat]} \BibitemShut
  {NoStop}%
\bibitem [{\citenamefont {Carrasco}\ \emph {et~al.}(2015)\citenamefont
  {Carrasco}, \citenamefont {Lubicz}, \citenamefont {Martinelli}, \citenamefont
  {Sachrajda}, \citenamefont {Tantalo}, \citenamefont {Tarantino},\ and\
  \citenamefont {Testa}}]{Carrasco:2015xwa}%
  \BibitemOpen
  \bibfield  {author} {\bibinfo {author} {\bibfnamefont {N.}~\bibnamefont
  {Carrasco}}, \bibinfo {author} {\bibfnamefont {V.}~\bibnamefont {Lubicz}},
  \bibinfo {author} {\bibfnamefont {G.}~\bibnamefont {Martinelli}}, \bibinfo
  {author} {\bibfnamefont {C.~T.}\ \bibnamefont {Sachrajda}}, \bibinfo {author}
  {\bibfnamefont {N.}~\bibnamefont {Tantalo}}, \bibinfo {author} {\bibfnamefont
  {C.}~\bibnamefont {Tarantino}}, \ and\ \bibinfo {author} {\bibfnamefont
  {M.}~\bibnamefont {Testa}},\ }\href {\doibase 10.1103/PhysRevD.91.074506}
  {\bibfield  {journal} {\bibinfo  {journal} {Phys. Rev.}\ }\textbf {\bibinfo
  {volume} {D91}},\ \bibinfo {pages} {074506} (\bibinfo {year} {2015})},\
  \Eprint {http://arxiv.org/abs/1502.00257} {arXiv:1502.00257 [hep-lat]}
  \BibitemShut {NoStop}%
\bibitem [{\citenamefont {Lubicz}\ \emph {et~al.}(2017)\citenamefont {Lubicz},
  \citenamefont {Martinelli}, \citenamefont {Sachrajda}, \citenamefont
  {Sanfilippo}, \citenamefont {Simula},\ and\ \citenamefont
  {Tantalo}}]{Lubicz:2016xro}%
  \BibitemOpen
  \bibfield  {author} {\bibinfo {author} {\bibfnamefont {V.}~\bibnamefont
  {Lubicz}}, \bibinfo {author} {\bibfnamefont {G.}~\bibnamefont {Martinelli}},
  \bibinfo {author} {\bibfnamefont {C.~T.}\ \bibnamefont {Sachrajda}}, \bibinfo
  {author} {\bibfnamefont {F.}~\bibnamefont {Sanfilippo}}, \bibinfo {author}
  {\bibfnamefont {S.}~\bibnamefont {Simula}}, \ and\ \bibinfo {author}
  {\bibfnamefont {N.}~\bibnamefont {Tantalo}},\ }\href {\doibase
  10.1103/PhysRevD.95.034504} {\bibfield  {journal} {\bibinfo  {journal} {Phys.
  Rev.}\ }\textbf {\bibinfo {volume} {D95}},\ \bibinfo {pages} {034504}
  (\bibinfo {year} {2017})},\ \Eprint {http://arxiv.org/abs/1611.08497}
  {arXiv:1611.08497 [hep-lat]} \BibitemShut {NoStop}%
\bibitem [{\citenamefont {Giusti}\ \emph {et~al.}(2018)\citenamefont {Giusti},
  \citenamefont {Lubicz}, \citenamefont {Martinelli}, \citenamefont
  {Sachrajda}, \citenamefont {Sanfilippo}, \citenamefont {Simula},
  \citenamefont {Tantalo},\ and\ \citenamefont {Tarantino}}]{Giusti:2017dwk}%
  \BibitemOpen
  \bibfield  {author} {\bibinfo {author} {\bibfnamefont {D.}~\bibnamefont
  {Giusti}}, \bibinfo {author} {\bibfnamefont {V.}~\bibnamefont {Lubicz}},
  \bibinfo {author} {\bibfnamefont {G.}~\bibnamefont {Martinelli}}, \bibinfo
  {author} {\bibfnamefont {C.~T.}\ \bibnamefont {Sachrajda}}, \bibinfo {author}
  {\bibfnamefont {F.}~\bibnamefont {Sanfilippo}}, \bibinfo {author}
  {\bibfnamefont {S.}~\bibnamefont {Simula}}, \bibinfo {author} {\bibfnamefont
  {N.}~\bibnamefont {Tantalo}}, \ and\ \bibinfo {author} {\bibfnamefont
  {C.}~\bibnamefont {Tarantino}},\ }\href {\doibase
  10.1103/PhysRevLett.120.072001} {\bibfield  {journal} {\bibinfo  {journal}
  {Phys. Rev. Lett.}\ }\textbf {\bibinfo {volume} {120}},\ \bibinfo {pages}
  {072001} (\bibinfo {year} {2018})},\ \Eprint
  {http://arxiv.org/abs/1711.06537} {arXiv:1711.06537 [hep-lat]} \BibitemShut
  {NoStop}%
\bibitem [{\citenamefont {Christ}\ and\ \citenamefont
  {Feng}(2018)}]{Christ:2017pze}%
  \BibitemOpen
  \bibfield  {author} {\bibinfo {author} {\bibfnamefont {N.}~\bibnamefont
  {Christ}}\ and\ \bibinfo {author} {\bibfnamefont {X.}~\bibnamefont {Feng}},\
  }\bibfield  {booktitle} {\emph {\bibinfo {booktitle} {{Proceedings, 35th
  International Symposium on Lattice Field Theory (Lattice 2017): Granada,
  Spain, June 18-24, 2017}}},\ }\href {\doibase 10.1051/epjconf/201817513016}
  {\bibfield  {journal} {\bibinfo  {journal} {EPJ Web Conf.}\ }\textbf
  {\bibinfo {volume} {175}},\ \bibinfo {pages} {13016} (\bibinfo {year}
  {2018})},\ \Eprint {http://arxiv.org/abs/1711.09339} {arXiv:1711.09339
  [hep-lat]} \BibitemShut {NoStop}%
\bibitem [{\citenamefont {Cai}\ and\ \citenamefont
  {Davoudi}(2018)}]{Cai:2018why}%
  \BibitemOpen
  \bibfield  {author} {\bibinfo {author} {\bibfnamefont {Y.}~\bibnamefont
  {Cai}}\ and\ \bibinfo {author} {\bibfnamefont {Z.}~\bibnamefont {Davoudi}},\
  }\href {\doibase 10.22323/1.334.0280} {\bibfield  {journal} {\bibinfo
  {journal} {PoS}\ }\textbf {\bibinfo {volume} {LATTICE2018}},\ \bibinfo
  {pages} {280} (\bibinfo {year} {2018})},\ \Eprint
  {http://arxiv.org/abs/1812.11015} {arXiv:1812.11015 [hep-lat]} \BibitemShut
  {NoStop}%
\bibitem [{\citenamefont {Christ}\ \emph
  {et~al.}(2015{\natexlab{b}})\citenamefont {Christ}, \citenamefont {Feng},
  \citenamefont {Martinelli},\ and\ \citenamefont
  {Sachrajda}}]{Christ:2015pwa}%
  \BibitemOpen
  \bibfield  {author} {\bibinfo {author} {\bibfnamefont {N.~H.}\ \bibnamefont
  {Christ}}, \bibinfo {author} {\bibfnamefont {X.}~\bibnamefont {Feng}},
  \bibinfo {author} {\bibfnamefont {G.}~\bibnamefont {Martinelli}}, \ and\
  \bibinfo {author} {\bibfnamefont {C.~T.}\ \bibnamefont {Sachrajda}},\ }\href
  {\doibase 10.1103/PhysRevD.91.114510} {\bibfield  {journal} {\bibinfo
  {journal} {Phys. Rev.}\ }\textbf {\bibinfo {volume} {D91}},\ \bibinfo {pages}
  {114510} (\bibinfo {year} {2015}{\natexlab{b}})},\ \Eprint
  {http://arxiv.org/abs/1504.01170} {arXiv:1504.01170 [hep-lat]} \BibitemShut
  {NoStop}%
\bibitem [{\citenamefont {Kim}\ \emph {et~al.}(2005)\citenamefont {Kim},
  \citenamefont {Sachrajda},\ and\ \citenamefont {Sharpe}}]{Kim:2005gf}%
  \BibitemOpen
  \bibfield  {author} {\bibinfo {author} {\bibfnamefont {C.}~\bibnamefont
  {Kim}}, \bibinfo {author} {\bibfnamefont {C.}~\bibnamefont {Sachrajda}}, \
  and\ \bibinfo {author} {\bibfnamefont {S.~R.}\ \bibnamefont {Sharpe}},\
  }\href {\doibase 10.1016/j.nuclphysb.2005.08.029} {\bibfield  {journal}
  {\bibinfo  {journal} {Nucl.Phys.}\ }\textbf {\bibinfo {volume} {B727}},\
  \bibinfo {pages} {218} (\bibinfo {year} {2005})},\ \Eprint
  {http://arxiv.org/abs/hep-lat/0507006} {arXiv:hep-lat/0507006 [hep-lat]}
  \BibitemShut {NoStop}%
\bibitem [{\citenamefont {Hasenfratz}\ and\ \citenamefont
  {Leutwyler}(1990)}]{Hasenfratz:1989pk}%
  \BibitemOpen
  \bibfield  {author} {\bibinfo {author} {\bibfnamefont {P.}~\bibnamefont
  {Hasenfratz}}\ and\ \bibinfo {author} {\bibfnamefont {H.}~\bibnamefont
  {Leutwyler}},\ }\href {\doibase 10.1016/0550-3213(90)90603-B} {\bibfield
  {journal} {\bibinfo  {journal} {Nucl. Phys. B}\ }\textbf {\bibinfo {volume}
  {343}},\ \bibinfo {pages} {241} (\bibinfo {year} {1990})}\BibitemShut
  {NoStop}%
\bibitem [{\citenamefont {Christ}\ \emph {et~al.}(2020)\citenamefont {Christ},
  \citenamefont {Feng}, \citenamefont {Jin},\ and\ \citenamefont
  {Sachrajda}}]{Feng:2020mmb}%
  \BibitemOpen
  \bibfield  {author} {\bibinfo {author} {\bibfnamefont {N.~H.}\ \bibnamefont
  {Christ}}, \bibinfo {author} {\bibfnamefont {X.}~\bibnamefont {Feng}},
  \bibinfo {author} {\bibfnamefont {L.-C.}\ \bibnamefont {Jin}}, \ and\
  \bibinfo {author} {\bibfnamefont {C.~T.}\ \bibnamefont {Sachrajda}},\ }\href
  {\doibase 10.22323/1.363.0259} {\bibfield  {journal} {\bibinfo  {journal}
  {PoS}\ }\textbf {\bibinfo {volume} {LATTICE2019}},\ \bibinfo {pages} {259}
  (\bibinfo {year} {2020})}\BibitemShut {NoStop}%
\end{thebibliography}%
\end{document}